\pdfoutput=1
\documentclass{article}

\usepackage{xcolor}
\usepackage{hyperref}
\usepackage{xurl}

\usepackage[accepted]{mlsys2025}

\makeatletter
\renewcommand{\mlsys@appearing}{\textit{Preprint.}  Copyright 2026 by the author(s).}
\renewcommand{\mlsysauthor}[2]{%
  \mbox{\bf #1}\@for\theaffil:=#2\do{\@pa{\theaffil}}\addtofullauthorlist{#1}}
\renewcommand{\@pa}[1]{%
  \ifcsname the@affil#1\endcsname\else
    \ifcsname @icmlsymbol#1\endcsname\else
      \stepcounter{@affiliationcounter}%
      \newcounter{@affil#1}%
      \setcounter{@affil#1}{\value{@affiliationcounter}}%
    \fi
  \fi}
\makeatother
\mlsystitlerunning{Taking Tokenization off the Host for Agentic LLM Serving}

\input{acr}
\newacronym{abi}{ABI}{application binary interface}
\newacronym{api}{API}{application programming interface}
\newacronym{bpe}{BPE}{byte-pair encoding}
\newacronym{dpu}{DPU}{Data Processing Unit}
\newacronym{kv}{KV}{key--value}
\newacronym{llm}{LLM}{large language model}
\newacronym{rag}{RAG}{retrieval-augmented generation}
\newacronym{slo}{SLO}{service-level objective}
\newacronym{sm}{SM}{streaming multiprocessor}
\newacronym{ttft}{TTFT}{time-to-first-token}

\newcommand{\Description}[1]{}

\usepackage{booktabs}
\expandafter\def\csname ver@fixltx2e.sty\endcsname{2016/12/29}
\usepackage{dblfloatfix} %
\usepackage{siunitx} %
\DeclareSIUnit\loc{lines\ of\ code} %
\DeclareSIUnit\LoC{LoC}

\usepackage{needspace}  %

\usepackage{enumitem}

\usepackage{microtype}
\usepackage{xspace}
\usepackage{tikz}
\usetikzlibrary{arrows.meta,decorations.pathreplacing}

\newcommand{\sys}{\textsc{GraphTok}\xspace}

\usepackage{pifont}
\newcommand{\stage}[1]{%
  \ifnum#1<11
    \raisebox{-0.1ex}{\ding{\numexpr181+#1\relax}}%
  \else
    \raisebox{-0.1ex}{\tikz[baseline=(n.base)]{%
      \node[circle,fill=black,text=white,inner sep=0.6pt,
            font=\tiny\bfseries] (n) {#1};}}%
  \fi}

\newcommand{\smartparagraph}[1]{\par\noindent\textbf{#1}\ \ignorespaces}

\usepackage{tcolorbox}
\newtcolorbox{takeawaybox}{colback=blue!5!white, colframe=blue!55!black,
  boxrule=0.4pt, leftrule=2.5pt, arc=1.5pt, left=5pt, right=5pt, top=3pt,
  bottom=3pt, fontupper=\small, before skip=6pt plus 2pt, after skip=6pt plus 2pt}
\newcommand{\takeaway}[1]{\begin{takeawaybox}\textbf{Takeaway.} #1\end{takeawaybox}}

\begin{document}

\twocolumn[
\mlsystitle{Taking Tokenization off the Host for Agentic LLM Serving}

\begin{mlsysauthorlist}
\mlsysauthor{Mohammad Siavashi}{kth},
\mlsysauthor{Pascal Schwinte}{kth},
\mlsysauthor{Ali Banaei}{kth},
\mlsysauthor{Gerald Q. Maguire Jr.}{kth},
\mlsysauthor{Marco Chiesa}{kth},
\mlsysauthor{Dejan Kosti\'{c}}{kth}
\end{mlsysauthorlist}
\mlsysaffiliation{kth}{KTH Royal Institute of Technology, Stockholm, Sweden}
\mlsyscorrespondingauthor{Mohammad Siavashi}{siavashi@kth.se}

\vskip 0.06in
\begin{center}
  {\small KTH Royal Institute of Technology, Stockholm, Sweden}
\end{center}

\vskip 0.2in

\begin{abstract}
Agentic workloads resend long contexts that barely change between calls. Prefix caching now skips most of their prefill, yet every request must still be tokenized in full before the cache can even be consulted. Tokenization, a CPU stage long treated as negligible, becomes the largest part of \gls{ttft}, and a frontend that tokenizes for a whole GPU fleet limits how many GPUs it can feed.
We present \sys, a GPU tokenizer that runs the entire encoding pipeline, from pre-tokenization through \gls{bpe} merging to token-ID assembly, as one replayable CUDA graph. \sys reproduces the reference token IDs bit-for-bit on GPT-2, Llama~3, Qwen~2.5, DeepSeek-V3, and Gemma~3, hands short inputs to a CPU engine where the GPU cannot win, and plugs into unmodified vLLM and NVIDIA Dynamo.
\sys encodes one million GPT-2 tokens in 1.7--3.2\,ms including data transfers, and on Llama~3 it is $4.7$--$6.0\times$ faster than gigatoken, the strongest CPU tokenizer. On TokTier's own single-request benchmark, it is $2.6$--$3.9\times$ faster than gigatoken and up to $18.1\times$ faster than TokTier. In vLLM with a cached prefix, it lowers mean \gls{ttft} by $2.1$--$2.4\times$.
The largest gain is capacity. Within a \qty{2}{\second} P99 target, one Dynamo frontend with \sys sustains 518 requests/s of 121k-token traffic, the busiest-hour load of 56 GPUs, against 296 with gigatoken and 18.5 with the default tokenizer. Across 121k to one million tokens, this is $1.75$--$2.5\times$ the load gigatoken sustains, with a P99 up to $64\times$ lower where gigatoken's tail breaks.

\end{abstract}
]

\section{Introduction}
\label{sec:intro}
With the rise of Large Language Model (LLM)-powered agents, a growing
fraction of LLM serving workloads comes from applications that invoke the
model repeatedly. A single coding task, for example, may involve many model
calls as an agent reasons, invokes a tool, observes its output, and continues
execution. Each call carries forward the accumulated conversation, code,
earlier tool calls, and retrieved data, while adding only a small amount of
new text. TraceLab's study of roughly 4,300 real coding-agent sessions
illustrates this asymmetry: the median model invocation contains about 119K
cached-prefix tokens but only 875 newly appended tokens and 214 output
tokens, and 95.7\% of input tokens are served from the prefix
cache~\cite{zhu2026tracelab}. Agentic workloads therefore combine long,
growing contexts with relatively small updates between consecutive model
calls.

\begin{figure}[t]
  \centering
  \includegraphics[width=\columnwidth]{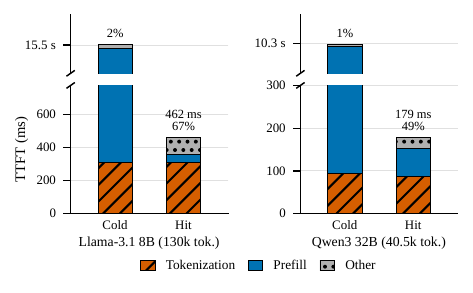}
  \caption{\Gls{ttft} breakdown for a cold cache and a prefix-cache hit in
  unmodified vLLM.  The y-axes are broken, and each model has its own scale.}
  \label{fig:teaser-breakdown}
  \Description{Two stacked-bar panels, one per model, each comparing a cold
  prompt against a prefix-cache hit. For Llama 3.1 8B (130k tokens), total
  TTFT falls from about 15.5 seconds to 462 milliseconds while tokenization's
  share rises from 2\% to 67\%. For Qwen3 32B (40.5k tokens), total TTFT
  falls from about 10.3 seconds to 179 milliseconds while tokenization's share
  rises from 1\% to 49\%. Orange hatched tokenization bars remain similar in
  absolute height as blue prefill bars shrink substantially, and gray dotted
  bars show other costs. Both vertical axes are broken and use different
  scales.}
\end{figure}

Modern LLM serving engines
\cite{kwon2023vllm,vllm2026prefixcache,zheng2024sglang,qin2025mooncake,nvidia2026dynamotokenizercache}
exploit this repetition through prefix caching. By reusing the
\gls{kv} state of previously processed prefix tokens, they avoid expensive
prefill computation over most of the repeated context and reduce
time-to-first-token (\gls{ttft}). One cost remains, however. The request must
still be tokenized before the \gls{kv} cache can be queried, because cache
lookup requires token IDs. Nor can the suffix simply be appended: appending
new text can change the resulting tokenization, so tokenizing only the new
suffix and concatenating it with a previously tokenized prefix is not safe in
general.

Tokenization nevertheless remains a CPU-side stage in common serving stacks,
even though model execution and the \gls{kv} cache reside on the GPU. This
division of work was historically reasonable because tokenization was
inexpensive compared with prefill. Prefix caching changes the balance: once
most prefill work is removed, the previously negligible tokenization stage
becomes a significant part of \gls{ttft}. Figure~\ref{fig:teaser-breakdown}
shows this effect in unmodified vLLM~0.27 with its CPU tokenizer on one H100
NVL. With a cold cache, tokenization accounts for only 2\% of median
\gls{ttft} for a 130k-token Llama~3.1~8B prompt and 1\% for a 40.5k-token
Qwen3~32B prompt. After a prefix-cache hit removes most of the prefill, these
shares rise to 67\% and 49\%, respectively, and tokenization becomes the
largest remaining component.

Three approaches can reduce this frontend cost. The first adds more CPU
tokenizer workers, improving aggregate throughput, but this is expensive,
does little for the latency of an individual long request, and competes for
CPU cores that retrieval, orchestration, networking, and tool execution also
need (Section~\ref{sec:motivation}). The second caches previously computed
token IDs. This helps for stable prefixes, but because appending text can change the tokenization, NVIDIA
Dynamo reuses token IDs only up to safe boundaries, such as completed chat
turns, and still tokenizes the remaining
text~\cite{nvidia2026dynamotokenizercache}. The third
moves tokenization to the GPU. BlockBPE and GPUTOK show that BPE merging can
be parallelized on the accelerator~\cite{you2025blockbpe,kadamba2026gputok},
but neither reproduces the complete reference-tokenizer pipeline. BlockBPE
changes the tokenizer semantics by removing the regular-expression
pre-tokenizer, and GPUTOK keeps part of the processing on the host but, in
dropping that pre-tokenizer as well, merges pairs that the reference keeps
separate (Section~\ref{sec:eval}).

In this paper, we present \sys, a GPU tokenizer that executes the complete
supported encoding path while preserving the ordered token IDs of the
reference tokenizer. \sys builds on the observation that both major stages of
\gls{bpe} tokenization contain more parallelism than their reference
implementations suggest. Pre-tokenization is usually implemented as a
sequential regular-expression scan, yet for the tokenizer families we
support, whether a new pre-token begins at a position can be determined from
local character classes together with compact run information. Accordingly,
\sys evaluates these boundaries in parallel across the input. \gls{bpe}
merging exposes a different form of parallelism: within a pre-token the merges
are ordered, but the threads of a warp cooperate to identify the next
highest-priority merge without changing the reference merge order.

Fast kernels alone do not make a fast tokenizer. A call that completes in a
few milliseconds still pays for repeated kernel launches, per-request
allocation, and host synchronization between stages. \sys therefore captures
the complete device-side encoding pipeline, from device-resident raw bytes to
the final token IDs, as a single CUDA graph. It separates a fixed graph
capacity from the actual work of each request: buffers and launch geometry
are provisioned in advance, and counters on the GPU record the live
occupancy. The same captured graph can therefore process any request up to
its capacity without intermediate allocation, synchronization, or host-side
dispatch.

We evaluate \sys on an NVIDIA H100 NVL with five production tokenizers,
GPT-2, Llama~3, Qwen~2.5, DeepSeek-V3, and Gemma~3. \sys produces the same
ordered token IDs as the corresponding reference tokenizer in every timed
cell of the evaluation campaign (Section~\ref{sec:eval-exactness}). Depending
on the tokenizer, the complete call tokenizes the largest English input, one
million GPT-2 tokens, in 1.7--3.2\,ms including transfers to and from the
GPU. On Llama~3 this is $4.7$--$6.0\times$ faster than gigatoken, the
strongest CPU baseline we measure. On the single-request protocol of
TokTier~\cite{zhang2026toktier}, a tokenizer developed concurrently and
independently, \sys is $2.6$--$3.9\times$ faster than
gigatoken with an empty cache and $2.0$--$18.1\times$ faster than the
released TokTier, from 100k to 4.4M characters.

The end-to-end benefit appears when prefix caching exposes the frontend.
With a cold cache, replacing vLLM's stock tokenizer improves mean \gls{ttft}
by only 2--4\%, whereas at 99\% reuse it improves mean \gls{ttft} by
$2.1$--$2.4\times$ for prompts between 32k and 128k tokens. For a single
uncontended request in one vLLM engine, however, fast CPU tokenizers come
within 4\% of \sys in mean \gls{ttft}, so the distinguishing benefit is frontend CPU and
capacity. As a separate process on a model's GPU, \sys does not measurably
slow inference, and under NVIDIA's Multi-Process Service (MPS) or a fine-grained scheduler it
runs on the same streaming multiprocessors (SMs) the inference leaves idle, overlapping it (Section~\ref{sec:eval}).

The capacity benefit materializes in NVIDIA Dynamo, where one frontend
tokenizes for a whole GPU fleet. There, \sys cuts the tokenization of a
121k-token request from \qty{235}{\milli\second} to
\qty{1.2}{\milli\second} (mean tokenize-stage time at one GPU's load). We report the sustained load in offered
requests/s and, in parentheses, as the number of GPUs whose traffic that
rate feeds, at 9.25 requests/s per GPU, BurstGPT's busiest hour
(Section~\ref{sec:eval-fleet}). Within a \qty{2}{\second} P99 target, the
default tokenizer lets a frontend sustain 18.5~requests/s (2 GPUs) of
121k-token agentic traffic and 4.6 (0.5) at 512k tokens. TokTier, with its
added-token boundary split, sustains 111 (12) and 18.5 (2), and \sys
sustains 518 (56) and 129.5 (14). This is $1.75\times$ the load the
frontend sustains with gigatoken at 121k--512k tokens and $2.5\times$ at one
million, and at the load where gigatoken's tail breaks, its P99 is up to
$64\times$ higher than that of \sys.

\smartparagraph{Challenges.}
Turning these observations into a tokenizer that a serving engine can use
raises three challenges.

\textbf{Preserving tokenizer semantics across different families.}
A production tokenizer does considerably more than execute the \gls{bpe}
merge loop (Section~\ref{sec:background}), and its preprocessing steps differ
between model families, from different regular-expression rules to
SentencePiece-style whitespace handling and byte fallback. A GPU
implementation therefore cannot simply accelerate one common merge kernel and
assume that the surrounding preprocessing is interchangeable. For every
supported configuration, \sys must parallelize the frontend while preserving
the pre-token boundaries, initial symbols, merge order, and final ordered
token IDs produced by the reference tokenizer.

\textbf{Handling input-dependent work inside a fixed CUDA graph.}
Tokenization is irregular. The number of pre-tokens, their lengths, the
number of merges within each one, and the final number of output tokens all
depend on the input. A straightforward GPU implementation would discover
these quantities at runtime, allocate buffers of matching size, return
intermediate counts to the host, and choose subsequent kernels based on those
counts. This is a poor fit for CUDA graphs, whose advantage comes from
replaying a previously captured execution without such host-side decisions.
Uncommon inputs that contain very long pre-tokens complicate the problem
further, because they must still be processed correctly rather than
rejected. \sys must therefore accommodate variable amounts of work within a
fixed captured execution.

\textbf{Making the complete encoding call fast.}
At millisecond-scale latencies, preprocessing left on the CPU, repeated
kernel launches, memory allocation, host--device synchronization, and output
assembly can cost as much as the computation itself. Existing serving engines
still expect host-visible token IDs for tasks such as request admission and
prefix-cache lookup, while future GPU-native consumers may prefer to keep the
result on the device. \sys must therefore accelerate the complete supported
encoding path rather than only the merge operation, and it must support both
host-returning and device-resident outputs. It must also account for the
fixed cost of GPU execution, which can make a CPU tokenizer preferable for
very short inputs.

\smartparagraph{Contributions.}
This paper addresses these three challenges with the following
contributions.

\smartparagraph{Exact parallel tokenization across heterogeneous frontends.}
\sys moves the frontend of each supported family, and not only \gls{bpe}
merging, onto the GPU. For each supported configuration, \sys parallelizes
these operations while preserving pre-token boundaries, merge
ordering, and the final ordered token IDs of the reference tokenizer
(Section~\ref{sec:design}).

\smartparagraph{Graph-resident execution of ragged tokenization.}
\sys separates the fixed capacity of a captured graph from the live
occupancy of each request, and an exact overflow path handles unusually long
pre-tokens without leaving the captured execution
(Section~\ref{sec:design}).

\smartparagraph{A complete serving-compatible GPU encoding path.}
\sys executes the supported frontend, \gls{bpe} merging, and token-ID
assembly on the GPU rather than accelerating only the merge kernels. It can
either return a conventional host-visible token sequence or expose the result
directly on the device. We integrate \sys into two serving systems and retain
the reference tokenizer as a fallback in both. It enters unmodified vLLM
through the tokenizer factory, and it serves NVIDIA Dynamo as a native
tokenizer backend of its Rust frontend, selected by one environment variable
(Section~\ref{sec:impl}). Our evaluation measures the complete
host-returning call, including transfers and output materialization, rather
than kernel time alone (Section~\ref{sec:eval}). \sys is open source and
available at \url{https://github.com/Msiavashi/GraphTok}.

\section{Motivation}
\label{sec:motivation}
Section~\ref{sec:intro} showed that prefix caching exposes tokenization on
the serving critical path. Before moving tokenization to the GPU, we ask
whether this operating regime occurs in real workloads and whether simpler
alternatives suffice: whether long prompts and high prefix reuse occur
together, whether token IDs can simply be reused instead of recomputed, and
whether additional CPU resources are enough to remove the remaining
latency.

\smartparagraph{Long contexts and high reuse occur together}
Tokenization becomes visible for requests that combine a long input with a
highly reusable prefix, and recent agent workloads exhibit this pattern
repeatedly. Agentix, for example, reports per-call prefix-cache hit rates above 90\%
within a single program, for chatbot inputs of up to about 150k
tokens~\cite{luo2026agentix}. KVFlow studies
repeated prompts in multi-agent workflows and finds substantial
opportunities for reusing their \gls{kv} state~\cite{pan2025kvflow}. Most
directly, the TraceLab coding-agent trace discussed in
Section~\ref{sec:intro} shows this combination at scale~\cite{zhu2026tracelab}.

The defining property is the asymmetry between the accumulated context and
the update between calls. An agent may carry tens or hundreds of thousands
of existing tokens from call to call while adding only a small tool result
or conversation turn. Prefix caching is particularly effective here, because
most model computation can be reused yet the frontend still receives a long
request. The high-reuse operating points in
Figure~\ref{fig:teaser-breakdown} therefore represent a regime observed in
real agent workloads, not an artificial corner case.

\takeaway{Long contexts and high prefix reuse occur together in real agent
workloads, and this is exactly the regime in which tokenization becomes
visible once prefill is reduced.}

\smartparagraph{Token reuse reduces, but does not eliminate, encoding}
Because much of an agent's input repeats across calls, an obvious
alternative is to cache the token IDs of the repeated text and tokenize only
the newly appended suffix. The difficulty is that \gls{bpe} tokenization
does not in general compose across an arbitrary text boundary. Appending
text can change the tokenization near the old endpoint, so tokenizing two
strings independently and concatenating their token sequences need not
produce the same result as tokenizing their concatenation. Exact incremental
treatments of \gls{bpe} therefore maintain state around the changing
boundary rather than encoding the suffix
independently~\cite{berglund-vandermerwe-2023-formalizing, jiang2026incrementalbpe}.

Production systems can still reuse token IDs when they know that a boundary
is safe. NVIDIA Dynamo, for example, caches prefixes that end at registered
special-token boundaries and encodes the remaining
suffix~\cite{nvidia2026dynamotokenizercache}. Such reuse is valuable and
should be used whenever possible, yet it does not remove the need for an
efficient encode path: reusable tokenization state may be unavailable, or
part of the request may still have to be encoded.

\takeaway{Token reuse and faster tokenization are complementary. Reuse avoids
redundant work for reusable regions and reduces the amount of text that
reaches the tokenizer, and \sys accelerates the exact encoding work that
remains.}

\smartparagraph{CPU resources improve capacity, but remain on the critical path}
Another alternative is to devote more CPU resources to tokenization. Running
more tokenizer workers raises frontend throughput by processing several
requests concurrently. In our measured setup, however, each Hugging Face or
tiktoken worker processes an individual request on one CPU core, so adding
workers does not reduce the latency of that request. Intra-request parallel
CPU tokenizers can reduce this latency further, and we compare against such
optimized implementations in Section~\ref{sec:eval}. Regardless of the CPU
implementation, tokenization still depends on host resources that other
parts of the serving stack also use.

Extra capacity is also not guaranteed. A centralized serving frontend such as
NVIDIA Dynamo's tokenizes every request for the whole GPU fleet before its
KV-aware router can choose a worker~\cite{nvidia2026dynamokvrouter}, so the
tokenizer's throughput bounds how many GPUs a single frontend can feed. In
Dynamo's default configuration, additional cores stop helping early. In a
saturation sweep on our 96-CPU host (included in the artifact), Hugging Face
encoding of 121k-token prompts plateaus at 45 requests/s with 90 CPUs busy,
because contention raises the CPU time of each request $12.5\times$.
Section~\ref{sec:eval-fleet} shows that, at the request rate of BurstGPT's
busiest hour, one Dynamo frontend with its default tokenizer keeps up with
the 121k-token agentic traffic of only two GPUs and with that of only half a
GPU at 512k tokens.

\begin{figure}
    \centering
    \includegraphics[width=0.9\linewidth]{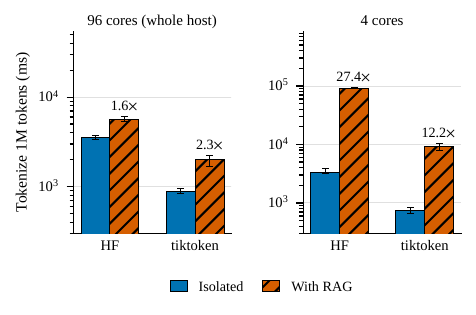}
    \caption{Latency of tokenizing 1M Llama~3 tokens in isolation and
    next to a RAG workload, on the full host and on four cores.  Bars are
    medians, and whiskers show the range.}
    \label{fig:cpu-contention}
\end{figure}

This dependence matters in agent systems, where retrieval, reranking,
orchestration, and other frontend work may execute alongside tokenization.
Figure~\ref{fig:cpu-contention} measures the effect by running tokenization
beside a retrieval-augmented generation workload that performs embedding,
\textsc{Faiss} retrieval, and cross-encoder reranking. With the full
96-thread host available, the co-located workload increases Hugging Face and
tiktoken latency by $1.6\times$ and $2.3\times$, respectively; under a
constrained four-core allocation, the same interference increases their
 latencies by $27.4\times$ and $12.2\times$ (medians of 15 loaded and 6
isolated trials). The two configurations show host-side tokenization at the
two limits a deployment can face, with and without a bounded CPU allocation,
and they bound how sensitive it is to competition for CPU resources.

\takeaway{More CPU workers add serving capacity only until contention sets
in, and optimized CPU tokenizers, while important baselines, still leave
tokenization a host-resident stage whose performance depends on the CPU
resources available. In a centralized frontend, that stage bounds how many
GPUs one host can feed.}

Taken together, these observations argue for accelerating, not replacing,
existing reuse mechanisms. Agent workloads create long, highly reused
contexts, token reuse removes some but not all encoding work, and the
remaining CPU encode is both latency-sensitive and exposed to host
contention. \sys therefore targets the complete encoding operation left after
reuse and moves it onto the GPU while preserving the reference token
sequence.

\section{Background}
\label{sec:background}
This section summarizes the tokenizer semantics and GPU execution constraints
that shape \sys. We focus on the properties the design depends on, namely
tokenizer-specific frontend processing, the ordering constraints of
\gls{bpe}, and the fixed execution structure imposed by CUDA graphs.

\smartparagraph{The tokenizer frontend is part of the model interface.}
\gls{bpe} merging is only one stage of a modern tokenizer. Before merging,
a tokenizer may recognize added or special tokens, normalize the input,
divide it into pre-tokens, and map each pre-token to an initial sequence of
symbols. These operations determine both the symbols presented to the merge
procedure and the boundaries across which merging is allowed, so replacing
or omitting any of them can change the final token-ID sequence.

Frontend semantics also differ across the tokenizer configurations that
\sys supports. GPT-2 and Llama~3 \cite{dubey2024llama3} both use byte-level
pipelines but apply different regular-expression splitting rules and
added-token configurations. Qwen~2.5 \cite{yang2024qwen25} follows the same
byte-level design, and DeepSeek-V3 \cite{deepseek2024tokenizer} applies
multiple splitting stages. Gemma~3 \cite{gemma2025tokenizer} instead uses a
SentencePiece-derived frontend that transforms whitespace, handles digits
differently, and falls back to bytes for Unicode symbols outside its
vocabulary. Supporting these configurations therefore requires
family-specific frontend logic, not a common pre-tokenizer with different
vocabularies.

\smartparagraph{BPE is parallel across pre-tokens but ordered within them.}
After frontend processing, the same ranked \gls{bpe} procedure applies to
every pre-token. Within a pre-token, the tokenizer repeatedly selects the
highest-priority eligible adjacent pair, merges it, and continues until no
merge rule applies. Different pre-tokens are independent and can be
processed in parallel. Within a single pre-token, however, merges are
ordered and data dependent, because each merge changes the neighboring pairs
that are eligible in the next step.

\smartparagraph{Exactness requires the complete ordered token sequence.}
Throughout the paper, an implementation is exact for a supported tokenizer
configuration and encode mode when it produces the same ordered token-ID
sequence as the corresponding reference tokenizer for the same input and
special-token settings. This definition covers added-token recognition,
normalization, pre-tokenization, initial-symbol construction or byte
fallback, and \gls{bpe} merge ordering. Matching only the token count or the
decoded text is insufficient, because the model consumes token IDs directly.
Request-level policies such as padding and truncation remain outside this
encoding contract.

\smartparagraph{CUDA graph replay favors fixed execution.}
CUDA graphs reduce launch overhead by recording a sequence of GPU operations
once and replaying the resulting execution graph. This makes them attractive
for a tokenizer whose useful computation completes within a few
milliseconds, but replay assumes a fixed execution topology, whereas the
quantities that size tokenization work are known only at runtime
(Section~\ref{sec:intro}).

\section{Design}
\label{sec:design}
\smartparagraph{Design requirements.}
\label{sec:design-overview}
The challenges in Section~\ref{sec:intro} translate into three requirements
that every part of the design must satisfy.
\begin{enumerate}[leftmargin=*, label=\textbf{D\arabic*}, ref=D\arabic*]
\itemsep0em
\item \textbf{Exact parallelism.} Parallel work must preserve pre-token
boundaries, initial symbols, merge order, and the order of the output token
IDs.
\item \textbf{Graph-resident raggedness.} Input-dependent pre-token counts
and output sizes must be resolved inside the captured graph, without
intermediate host control or allocation.
\item \textbf{Complete GPU path.} The accelerated path must include the
frontend, the merges, and the assembly of the result, so that no remaining
host work becomes the latency bottleneck.
\end{enumerate}

\begin{figure}[t]
  \centering
  \includegraphics[width=0.7\columnwidth]{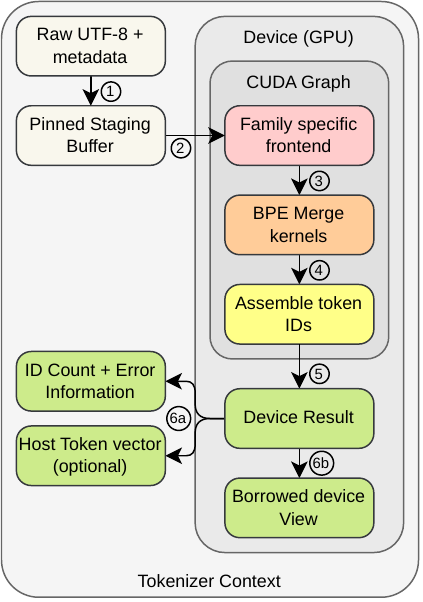}
  \caption{\sys encode dataflow.  Numbered arrows are handoffs between
  stages, and 6a and 6b are the host and device return paths.}
  \label{fig:design-dataflow}
  \Description{A tokenizer context contains a host staging area and a GPU
  dataflow. Boxes name computations and numbered arrows name the data handoffs
  between them. Arrow 1 moves raw UTF-8 text and request metadata into a pinned
  staging buffer, and arrow 2 transfers them to the GPU. Inside one CUDA graph,
  the family-specific frontend produces initial symbols and boundaries, which
  arrow 3 passes to the BPE kernels. Arrow 4 passes merged per-pre-token results
  to token-ID assembly, and arrow 5 writes the ordered IDs into a device result
  buffer. Arrow 6a returns the output count, error information, and optionally
  a host token vector. Arrow 6b exposes a borrowed view of the same device
  result and can accompany the metadata-only form of 6a.}
\end{figure}

\smartparagraph{Overview.}
Figure~\ref{fig:design-dataflow} shows how a request flows through \sys.
Raw input is staged in pinned memory~\stage{1} and transferred to the
device~\stage{2}, and the frontend passes initial symbols and pre-token
boundaries to \gls{bpe}~\stage{3}. The central interface is the split
between a family-specific frontend and a shared backend. Each supported
tokenizer family transforms raw text into ordered initial symbols and
pre-token boundaries, and the shared backend then assigns the independent
pre-tokens to merge work, preserves their document order, and compacts the
results into the final token-ID sequence. For batched calls, the offsets of
the individual documents act as additional hard boundaries in the same
representation, so neither pre-tokenization nor merging can cross from one
document into the next.

Three operations stay outside the graph. These are recognizing which
configuration a tokenizer file describes, request-level policies such as
truncation, and the Unicode NFC normalization that Qwen's tokenizer applies
before pre-tokenization. The last is a host pass with a quick check that
returns ASCII and already-normalized text unchanged, so on such input the
graph still receives the raw bytes. We also chose to reject tokenizer
structures that the frontends do not support rather than assign them
approximate semantics, since an approximate tokenizer would violate the
exactness contract silently.

\smartparagraph{Parallel tokenizer frontends.}
\label{sec:design-pretok}
The first requirement, exact parallelism, is hardest to meet in the
frontend, because the reference pre-tokenizer is a regular-expression
matcher that advances from one match to the next and thereby serializes the
whole step. None of the three obvious ways around this is satisfactory.
Keeping the matcher on the host and moving only the merges to the GPU would
leave the frontend on the critical path and exposed to host contention
(Section~\ref{sec:motivation}); a general regular-expression engine for the
GPU would solve a larger problem than we have, since the supported patterns
are few and known in advance; and dropping the regular expression altogether,
as BlockBPE and GPUTOK do, changes the tokens.

\sys takes a fourth route. For each supported pattern, it derives a boundary
predicate that decides whether a pre-token begins at a given position. At
every code-point position, a GPU thread evaluates this predicate from local
Unicode classes and, where a rule needs it, from run metadata. Prefix scans
supply the context that individual threads lack, such as the offset of a
character within a capped digit run, and also turn the resulting boundary
flags into ordered pre-token indices~\cite{merrill2016prefixscan}. Llama~3's
rule \verb|\p{N}{1,3}|, for example, cuts a digit run into groups of at most
three from its left end, and no other alternative of the pattern attaches a
neighbor to a digit. A digit therefore begins a pre-token exactly when it
begins its run or its offset within the run is a multiple of three. A
segmented scan over the digit flags supplies that offset, so in
\texttt{x12345} the predicate fires at \texttt{1} and \texttt{4} and yields
\texttt{123} and \texttt{45}, exactly as the left-to-right matcher does.
 This formulation covers the distinct GPT-2, Llama~3, Qwen~2.5, and
three-pass DeepSeek-V3 rules without a serial regex cursor; each supported
family contributes one predicate, which is the price of the
exactness-first design of Section~\ref{sec:design-overview}.

Two further mechanisms complete the frontend. Added vocabulary entries take
precedence over ordinary pre-tokenization, so a device lookup first
identifies supported literals, emits their IDs directly, and passes only
the unmatched regions to the family predicate. Gemma~3, whose pipeline is
not based on a regular expression at all, has a separate frontend that maps
spaces to U+2581, emits vocabulary IDs for known Unicode symbols, and uses
byte fallback otherwise. Despite these differences, every frontend produces
the same ordered symbol-and-boundary representation, which is what allows
the merge backend to be shared.

\smartparagraph{Warp-cooperative BPE.}
\label{sec:design-merge}
The merges pose the opposite problem. Parallelism across pre-tokens comes
for free, because \gls{bpe} never crosses a pre-token boundary; within one
pre-token, however, the merges are ordered and data dependent, and the
reference order must be preserved exactly. \sys therefore parallelizes across
pre-tokens first and, within a pre-token, parallelizes the search for the
next merge. It applies several merges of one pre-token at once only where it
can show that sequential \gls{bpe} would apply exactly those merges
consecutively, which matters for the rare very long pre-token described
below.

At the frontend/backend boundary, pre-tokens are bucketed by their number of
initial symbols, and each bucket takes one of three paths. A warp handles
each pre-token of at most 64 symbols, with two independent warps per block.
For up to 32 symbols each lane owns one symbol, and for 33 to 64 each lane
owns two. Where the reference skips the merges for a pre-token that is
itself a vocabulary entry (Llama~3's \texttt{ignore\_merges}), one lane
first looks up the whole pre-token and emits its ID directly. In every
iteration, the lanes look up the merge ranks of their adjacent pairs, pack
each $(\mathit{rank},\mathit{position})$ key into one word, and reduce the
keys with the warp's hardware minimum (\texttt{\_\_reduce\_min\_sync} on
sm\_80 and later, shuffles otherwise). The lowest rank wins, and the
position component resolves equal-rank occurrences leftmost first, which is
exactly the reference merge order~\cite{zouhar-etal-2023-formal,tiktoken}.
Each probe also returns the merged ID, so the merge itself needs no second
lookup. After a merge, only the winning pair and its new neighbors need
updating, which keeps each iteration cheap.

Pre-tokens longer than 64 symbols are rare, but they cannot simply be
rejected, because the correctness contract covers every input. Allocating
extra memory or retrying on the host would both break graph capture, so
\sys routes such pre-tokens to an overflow path spread over every
SM. Up to 512 symbols, which covers every overflow pre-token of our three
benchmark corpora at every size, a warp keeps a tournament tree in shared
memory that repeatedly selects the lowest
$(\mathit{rank},\mathit{position})$ pair and thereby preserves the merge
order of the fast path.

Longer pre-tokens are almost always runs of one repeated character, for
which sequential \gls{bpe} applies the same merge many times in a row, one
merge per step. Dedicated blocks of the same launch therefore apply these
merges in rounds. A round computes the minimum rank $r$ with a block-wide
reduction and selects every other pair of each maximal chain of adjacent
rank-$r$ pairs with a segmented scan. For each selected merge, it evaluates
the pairs that the merge forms with the neighbors it has at the moment
sequential \gls{bpe} applies it: the result of the selection two positions
to its left, or the original left symbol, and the original right symbol.
The round then applies the selected merges up to and including the first
that forms a pair of rank at most $r$.

These rounds reproduce the sequential order. Because ranks are unique,
every pair of rank $r$ is an occurrence of the same merge, and sequential
\gls{bpe} takes these occurrences from left to right. Within a chain it
takes every other pair, because each merge consumes the left symbol of the
pair to its right. It continues this way as long as no merge creates a pair
of rank at most $r$, since it would take such a pair before the remaining
rank-$r$ pairs to its right. Every merge a round applies is thus a step of
the sequential execution, in the same order, and the next round restarts
from the true minimum.

A run of one character finishes in a handful of rounds. A round that
applies fewer than $\max(2, n/1024)$ merges to an $n$-symbol pre-token hands
the remainder to the tree, so non-repetitive text pays for at most one
unproductive round before the tree takes over. All of this storage
is provisioned together with the graph, so an unusual pre-token costs time
but never an allocation. The three paths run as parallel branches of the
graph, and once they finish, handoff~\stage{4} passes the per-pre-token results
to assembly, which scans their output lengths, restores document order, and
writes the device result through handoff~\stage{5} in one fused kernel. For
capacities of at most 4\,KiB, a single thread block performs the whole
frontend and another performs the scan and assembly, so a small graph has
few nodes.

\smartparagraph{Graph-resident variable-length execution.}
\label{sec:design-graph}
The second requirement is to run this pipeline as one CUDA graph even though
every quantity in it depends on the input. A graph can be updated between
replays, but updating it for each request would put the host back into the
loop for every encode. \sys instead reuses one instantiated topology for
each capacity~\cite{nvidia2026cudagraphs}. The mechanism that makes this
possible is a separation, inside a tokenizer context, between fixed
\emph{capacity} and live \emph{occupancy}. Buffers and launch grids are
sized for a bounded input size and pre-token count, while device-resident
counters and masks identify how much of that capacity a particular replay
actually uses. All allocation and all temporary-storage sizing happen
before capture, and no intermediate result ever returns to the host to
select subsequent work.

In steady state, the host consequently does very little. It copies the raw
text and a small header of lengths and document offsets to the device and
launches the graph once. Handoff~6a always returns the output count and
status, which the host reads from mapped pinned memory without a copy, and
it copies the compact token range only when host IDs are requested.
Handoff~6b exposes the same device buffer directly, so a caller can receive
only metadata on the host and continue with the tokens on the device.
Because a graph's capacity adds fixed and inactive work to every replay on
top of the input-dependent work, \sys keeps a small ladder of graphs, one
per power-of-two input size, and replays the class of each request. The one
departure from this path is a size class seen for the first time: \sys
captures that class's graph once, and later requests of the same class reuse
it.

\smartparagraph{Serving interface.}
The third requirement, a complete path, raises the question of where the
token IDs should end up, and \sys supports both answers. The evaluated vLLM
adapter uses the host-compatible output path, because vLLM consumes the
token IDs on the host during admission to count prompt tokens, to construct
prefix-cache block hashes, and to apply request policy, so a
device-resident result would have to be copied back regardless. \sys thus
replaces plain-text encoding while preserving the tokenizer interface that
surrounds it. The device path (handoff~6b) remains available for downstream
GPU consumers that keep only the count and status metadata on the host, and
such consumers already exist. CPU-free inference systems such as
Blink~\cite{siavashi2026blink} move the entire serving stack, including
tokenization, onto the accelerator so that token IDs never cross the
host--device boundary. For such a system, \sys's device-resident output is
the only output that preserves the benefit of graph-rate execution, because
a host-returning call would reintroduce exactly the transfer and
synchronization that a CPU-free path is designed to remove.

\section{Implementation}
\label{sec:impl}
\smartparagraph{Runtime and asset loading.}
\sys is implemented in C++17 and CUDA behind a C \gls{abi}, and a
command-line tool and a pybind11 extension share the same tokenizer context.
Because the exactness contract is defined per configuration rather than per
model, the loader identifies its input structurally: it parses Hugging Face
\texttt{tokenizer.json} files and recognizes a supported family by its
serialized normalizer, pre-tokenizer, \gls{bpe} options, and added-token
behavior, not by model name, and it rejects outright any configuration that
contains unsupported semantics. \sys supports GPT-2, Llama~3, Qwen~2.5
(whose configuration Qwen~3 shares), DeepSeek-V3, and Gemma~3.
Single-vocabulary builds compile only the frontend and table policy they
need, whereas a combined build dispatches at runtime among the families it
includes. For Qwen, the loader also applies NFC normalization on the host
before the encode, using utf8proc behind the quick check described in
Section~\ref{sec:design-overview}, and it rejects added tokens that
normalization could alter. Counting code lines without blanks and comments,
the library is about 11k lines of C++ and CUDA, of which about 4.8k are the
GPU kernels and graph capture, plus a 100-line Rust crate that links
gigatoken's engine for the CPU route; the serving integrations below are
small by comparison.

\smartparagraph{Device representation.}
Every warp lane looks up the merge rules in every merge iteration, so their
device layout matters. At context construction, the host resolves the
operands of every merge to token IDs and builds an open-addressed device
table that is at most half full. GPT-2 fits a merge into a packed 64-bit
entry containing the two input IDs, the output ID, and the rank, while the
larger configurations use separate 64-bit key and value arrays with 32-bit
IDs and ranks. Alongside the table, the loader uploads the byte-to-ID map,
the added-token trie, and any family-specific direct-lookup metadata, and
each context places compact Unicode-class tables in constant memory.

\smartparagraph{Captured execution.}
Each tokenizer context owns a nonblocking CUDA stream, two auxiliary streams
that fork and join the three merge paths during capture, pinned and device
buffers, scan workspace, and one executable graph for its capacity. Capture
must be prepared carefully, because anything that happens lazily inside it
would break the graph. Construction therefore first queries CUB for its
temporary-storage requirements and runs a warmup so that module loading does
not occur during capture; only then does it capture the frontend, the three
merge paths, and result assembly with \texttt{cudaStreamBeginCapture}.
The kernels are sized for the capacity and consult device counts or
empty-row sentinels to skip inactive work.

Handoff~6b exposes the internal result buffer through
\texttt{\_\_cuda\_array\_interface\_\_} as a borrowed \texttt{int32} tensor,
and the next encode invalidates that view unless the caller copies it. Above
the context sits a ladder of at most eight contexts, one per power-of-two
size class with a floor of 1\,KiB (4\,KiB in the C library). An encode
replays the context of its own class and creates it on first use; once the
ladder is full, it reuses the tightest class that fits. The Python
extension, the C library, and the dispatcher each keep such a ladder.

\smartparagraph{Batch entry point.}
For the byte-level tokenizers, a single call to the batch entry point,
\texttt{encode\_batch()} in Python and \texttt{gtok\_encode\_batch()} in
the C library, concatenates the documents of a batch in one pinned input
buffer and passes their byte offsets to the graph in a small header. The
offsets keep both pre-tokenization and merging document-local, so two
host-to-device copies and one graph replay produce a concatenated token
array together with per-document output offsets. The single-document entry
point is simply a batch of one. The current Gemma frontend does not support
batching.

\smartparagraph{CPU route and dispatcher.}
Below a measured input size, the graph's fixed cost exceeds the whole
encode on one CPU core, so the library routes such inputs to a CPU engine.
This engine is gigatoken (MIT-licensed), linked statically through a small
Rust crate with its own exported symbols, with our earlier host encoder as
the fallback when the toolchain is absent. At construction, gigatoken must
agree with the host encoder on a probe of pre-tokenizer edge cases and added
tokens, or the CPU route falls back to the host encoder; DeepSeek-V3, which
has no host encoder, skips the probe. One router serves the Python and C
interfaces. It applies measured crossover limits and, near them, compares
the costs it observes on each route (Section~\ref{sec:eval-cpu-route}). For
many concurrent callers, a thread-safe dispatcher with an adaptive policy
runs short inputs on a
small pool of CPU engines when one is free and otherwise combines waiting
requests into one graph replay, launched when the previous one completes.
The vLLM adapter and the Dynamo backend can use this dispatcher in place of
per-call contexts.

\smartparagraph{Interface boundaries.}
The graph implements bare encoding and nothing else; everything the design
leaves on the host (Section~\ref{sec:design-overview}) stays there, and the
complete-call timings include Qwen's host NFC pass. Drawing the boundary here
lets a serving integration substitute the encoding step without replacing the
complete tokenizer object.

\smartparagraph{vLLM adapter.}
The integration is a Python adapter that is loaded before vLLM constructs
its tokenizer. It wraps the renderer's tokenizer factory and the
asynchronous microbatch tokenizer without modifying vLLM source, and it
activates only for a supported family (Llama~3, Qwen~2.5, and Qwen~3 in the
serving evaluations). The adapter loads \sys's C library, accelerates
\texttt{encode} and batched string calls, and forwards every other attribute
and method to the reference Hugging Face object. Single prompts call
\texttt{gtok\_encode}, batched calls use \texttt{gtok\_encode\_batch}, and a
small C extension builds the Python list of IDs directly from the library's
output.  The adapter is about 310 lines of Python plus a 100-line C
extension.

Two details of how vLLM uses its tokenizer required care. First, vLLM passes
\texttt{truncation=True} as a length guard, so the adapter performs the GPU
encode and slices only outputs that exceed \texttt{max\_length}. Second, the
reference tokenizer may add prefix or suffix tokens that the bare graph does
not, so a probe at construction reconciles the two. An unsupported family
or an irreconcilable probe stops the server from starting rather than
silently serving reference tokens, and ordinary runtime failures fall back
to the reference method. Correctness is also gated outside the adapter.
Before every serving run, the harness checks that the adapter's IDs equal
the reference tokenizer's on the run's prompts and refuses to report the run
otherwise, and it checks a trace counter to confirm that the accelerated
requests actually took the \sys path.

\smartparagraph{Dynamo backend.}
NVIDIA Dynamo's frontend is written in Rust and tokenizes through an encoder
interface in its tokenizer crate that already selects among backends by an
environment variable. We add \sys as a native backend, selected with
\texttt{DYN\_TOKENIZER=graphtok}, in about 200 lines of Rust plus about 60
changed lines in Dynamo's tokenizer switch and command line. The backend loads
\sys's C library at runtime, so Dynamo neither links against CUDA nor needs
a GPU unless the backend is selected. A graph context serves one caller at
a time, so by default the backend lends each concurrent frontend worker a
context from a fixed pool on one GPU. With
\texttt{DYN\_GRAPHTOK\_DISPATCH=1}, which the fleet evaluation uses, all
workers instead share one dispatcher with the adaptive policy, one CPU
engine, and batches of at most 8\,MiB and 256 documents, so that concurrent
prompts arriving together cost one graph replay. In both modes the backend
keeps the ladder of graphs described above. Decoding is
delegated to the reference tokenizer. As in the vLLM adapter, construction
checks that a chat-templated probe produces the reference token IDs and
refuses to load otherwise, and a failed call falls back to the reference
encoder.

\section{Evaluation}
\label{sec:eval}
\smartparagraph{Evaluation questions.}
The evaluation answers three questions about the performance of \sys.
\textbf{Q1:} How does complete GPU tokenization compare with production CPU
and GPU tokenizers across input lengths, text distributions, and batch sizes,
including on TokTier's own benchmark, and what do the CPU route and the
dispatcher add?
\textbf{Q2:} Does the tokenizer-level improvement reduce serving latency once
prefix reuse has removed model work from the request path, and how large a
GPU fleet can one serving frontend then feed, including on a host shared with
coding agents?
\textbf{Q3:} How much does graph replay save, and where does the time go
inside the complete tokenizer?
 Exactness is a precondition of the measurement, so
Section~\ref{sec:eval-exactness} establishes it before any timing, and we
also evaluate batching as a secondary capability.

\subsection{Experimental setup}

All experiments run on one NVIDIA H100 NVL (compute capability 9.0) with
driver 595.71.05 and CUDA 12.6, hosted by two Intel Xeon Gold 6336Y
processors (96 hardware threads) with \qty{256}{\giga\byte} of DRAM and
the CPU frequency governor set to \texttt{performance}.  The standalone and
batch comparisons are one measurement each, taken with a single build of
\sys and with every system called through its own Python interface in one
process, so each timing covers what a library caller pays.  The stage
profile, the graph ablation, and the serving experiments are separate runs
on the same GPU.  The CPU baselines are tiktoken~0.12.0, Hugging Face
tokenizers~0.22.2, gigatoken~0.10.0, and TokTier~0.2.9 in two modes.  In
the standalone and batch comparisons, every library runs with its default
thread pool and no affinity restriction and may therefore use all 96
hardware threads, the most favourable setting for engines that parallelize
internally.  The serving experiments instead pin the server to the CPU
socket the GPU is attached to (Section~\ref{sec:eval-burst}).  Nothing else
ran on the host during the standalone and batch measurements.

We use three corpora with different character distributions, namely English
books from Project Gutenberg~\cite{gutenberg}, CPython source and technical
prose~\cite{cpython}, and multilingual chat messages spanning 35 languages
from OpenAssistant Conversations~\cite{kopf2023openassistant}.  So that every
tokenizer sees the same text, each corpus provides ten fixed input files
whose sizes are set in GPT-2 tokens, from 128 to one million.  Because
vocabularies split text differently, the same file yields a different token
count under each tokenizer.  The largest books file (\qty{3.8}{\mega\byte})
is exactly one million GPT-2 tokens and 0.91--0.92 million Llama~3,
Qwen~2.5, DeepSeek-V3, and Gemma~3 tokens, and the largest code and
multilingual files are 0.72 and 0.63 million Llama~3 tokens.  Figures plot
each point at the token count its tokenizer actually produced.  Every cell
runs three untimed and twenty timed calls.  We report the median, and the
archive keeps every sample together with its 95th and 99th percentiles.
Cells are measured one at a time, cycling through all systems, so drift
affects every system alike.  The stage profile and the graph-execution
ablation use their own repetition schedules, described in
Section~\ref{sec:eval-execution}.

Vocabulary loading, allocation, and CUDA graph capture all happen before
measurement.  We distinguish two timings.  Graph-only timing places CUDA
events around one replay with input and output already on the device.
Complete-call timing is the wall time of the library call a user makes,
which returns token IDs in host memory.  For \sys this call is the library
as shipped.  Inputs below a measured crossover take its CPU route
(Section~\ref{sec:eval-cpu-route}), and the rest take the graph, including
raw-text handling, both transfers, and output materialization.  The campaign
archive contains the raw samples, reference-comparison records, environment
and binary hashes, and figure-generation scripts.

\begin{figure*}[t]
  \centering
  \includegraphics[width=\textwidth]{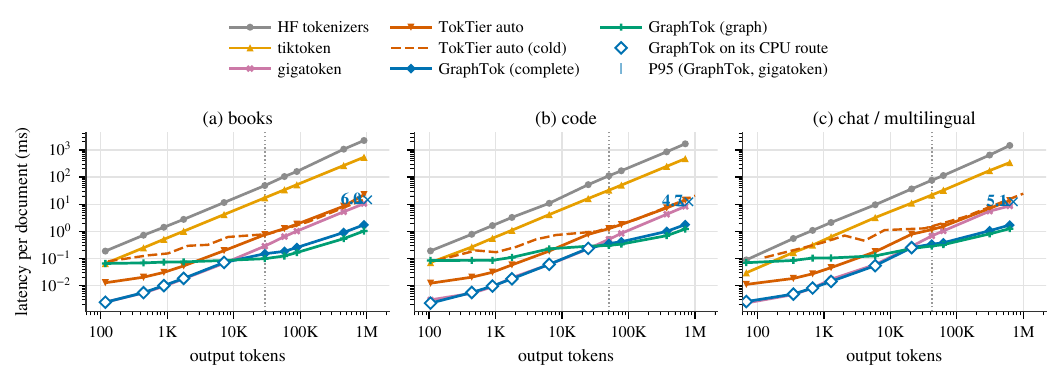}
  \caption{Complete tokenization latency, one document per call, on the
  Llama~3 vocabulary (log--log, medians of 20 calls, P95 whiskers on \sys
  and gigatoken).  Open markers are inputs that \sys serves on its CPU
  route, where it matches gigatoken.  The dotted line marks the first input
  its GPU graph takes over (the 32k file on books, the 64k file on code and
  chat).  TokTier is shown warm (solid) and cold on never-seen text
  (dashed).  The annotated gap is \sys's lead over gigatoken and warm
  TokTier on the largest input.}
  \label{fig:standalone-cpu-gpu}
   \Description{Three log--log panels compare single-document tokenization
    latency from 68 to 913k output tokens on the books, code, and
    multilingual chat corpora.  \sys coincides with gigatoken on short inputs
    (open markers, CPU route) and becomes the lowest curve from the 32k--64k
    input onward; the green curve is \sys's graph-only path.}
\end{figure*}

\begin{figure*}[t]
  \centering
  \includegraphics[width=\textwidth]{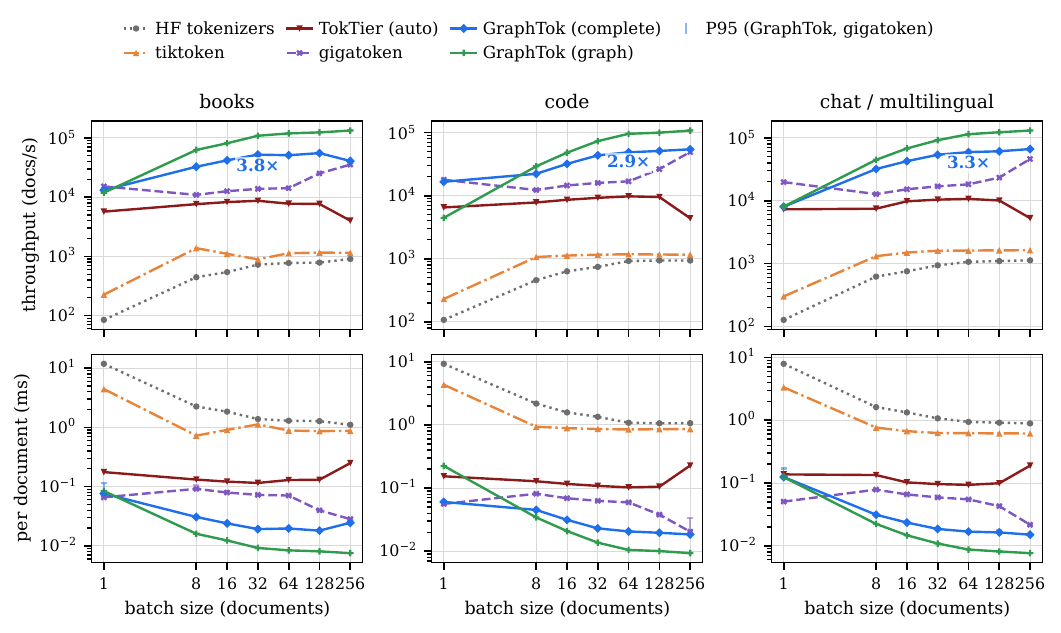}
  \caption{Aggregate throughput (top) and amortized time per document
  (bottom) for 8k-token Llama~3 documents (log--log, medians of 20 calls,
  P95 whiskers on \sys and gigatoken).  \sys trails gigatoken at batch~1,
  where the call is a single-document call that the router serves on its
  CPU engine for books and code, while the graph-only series forces the GPU
  graph, which one document leaves mostly idle.  From batch~8 on, \sys leads
  every system.  The annotated number is its peak lead over gigatoken.}
  \label{fig:batch-scaling}
  \Description{Six log--log panels in two rows show throughput and amortized
  time per document for the books, code, and multilingual chat corpora as
  batch size grows from 1 to 256.  GraphTok's complete and graph-only paths
  start near or below gigatoken at batch 1 and become the highest-throughput
  implementations from batch 8 onward, while the CPU baselines flatten.}
\end{figure*}

\begin{figure*}[t]
  \centering
  \includegraphics[width=\textwidth]{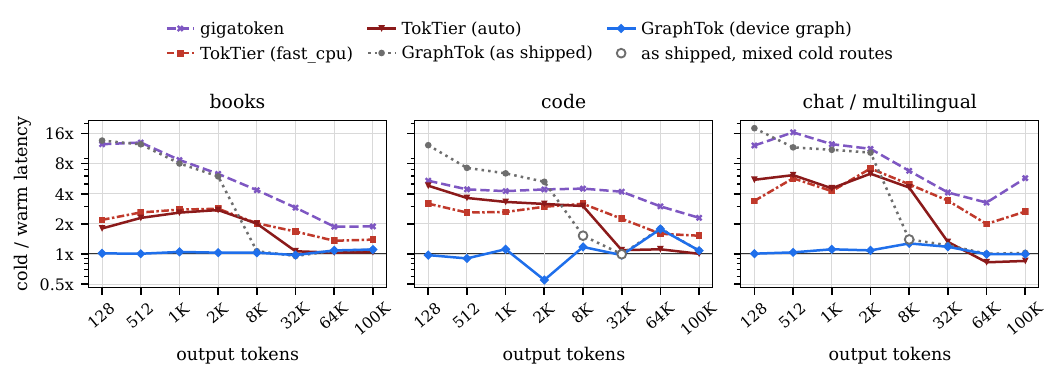}
   \caption{Ratio of the latency on never-repeated documents to the latency of
   re-encoding one document, one document per call (Llama~3, log $y$, medians
   of 20 calls).  A ratio of $1\times$ means no penalty.  The CPU engines'
   ratios are lower bounds, because most of their cells re-read text an
   earlier cell had seen, so the plotted value is the first-sight penalty
   divided by a warm-cache latency.  The open markers are \sys cells whose
   cold arm mixed CPU and GPU routes, and the single point below one is a
   warm-side variance effect, not a cold-arm speedup.}
  \label{fig:cold-warm}
   \Description{Three log-scale panels plot the ratio of cold to warm
    tokenization latency against input size, 128 to 100k tokens, for the
    books, code, and multilingual chat corpora.  GraphTok's device graph
    (blue) stays near one at every size, while the CPU engines sit above it;
    GraphTok as shipped (gray) matches the device graph from 8k onward, with a
    few mixed-route points drawn open.}
\end{figure*}

\begin{figure}[t]
  \centering
  \includegraphics[width=\columnwidth]{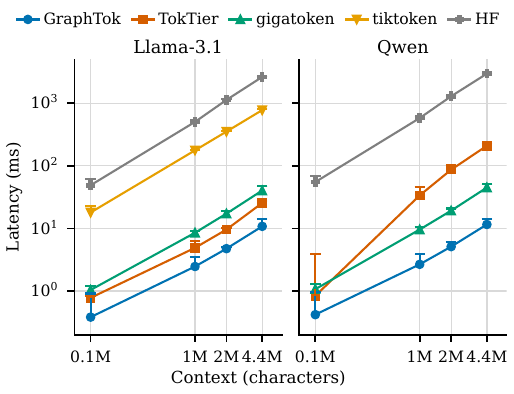}
  \caption{Complete tokenization of one fresh coding-agent context per call,
  following TokTier's single-request protocol (SWE-smith
  \cite{yang2025swesmith} contexts, 20 per size, CPU engines
  single-threaded, gigatoken with an empty cache, log--log).  Markers are
  medians and whiskers reach the P99.  Every engine returns a Python list
  and matches Hugging Face on every sample.  tiktoken, which lacks Qwen's NFC
  normalization, is shown for Llama~3.1 only.}
  \label{fig:single-request}
   \Description{Two log--log panels, Llama 3.1 and Qwen, plot median
    single-request latency with whiskers up to the 99th percentile for
    contexts of 100 thousand, 1, 2 and 4.4 million characters.  GraphTok is the
    lowest line in both panels and its P99 whiskers are the lowest from 1 million
    characters up; tiktoken and Hugging Face are one to two orders of magnitude
    slower.}
\end{figure}

 \subsection{Does \sys produce the reference token IDs?}
\label{sec:eval-exactness}

Exactness is the precondition for every other result, so we establish it
first.  For every measured input, we compare \sys's complete ordered
token-ID sequence with the output of a reference tokenizer on the same text.
The reference for GPT-2 is tiktoken~0.12.0, and the reference for Llama~3,
Qwen~2.5, DeepSeek-V3, and Gemma~3 is Hugging Face tokenizers~0.22.2, loaded
from the same serialized tokenizer assets as \sys.  The comparison uses bare
encoding, with automatic BOS/EOS insertion disabled while registered
added-token literals that already appear in the text remain recognized,
which keeps the special-token settings aligned between the implementations.

We check \sys's output against the reference before timing in two
independent sets.  The first is a campaign spanning five vocabularies, three
corpora, ten input sizes, and batch sizes from 1 to 256, and every cell
matches.  The campaign drives \sys's standalone binary, which
has no CPU route, so every cell exercises the GPU path, including the short
inputs that the library would otherwise serve on the CPU.  A batch of $k$
repeats one document $k$ times in one replay, so every document seam is
checked; Gemma~3, which has no batch path, encodes the same documents one
call at a time.  The largest batches are skipped where their input exceeds
the workspace cap, and Gemma~3, whose larger vocabulary needs more
workspace, runs out of memory on a few more.  The build of
22--24~September and the final build give the same matching cells and the
same skips.

The second set consists of the standalone, batch, and
vocabulary measurements reported below, taken afterwards on the final
build.  Every system's output, \sys's on whichever route it took, is
checked against the reference before timing, and all match.  The same
check decides which CPU baselines we report: only engines that reproduce
the reference sequence on the measured cells enter the latency comparison,
so the comparisons below are among exact engines.  tiktoken lacks the NFC
normalization that DeepSeek-V3 and Gemma~3 require, so it is listed as
inapplicable to those vocabs rather than as a failure or a zero.

Because the corpus sweep exercises common text, targeted tests cover the
cases it would miss, including empty and short text, whitespace and digit
runs, Unicode text, and added-token literals adjacent to ordinary text.
Long pre-tokens exercise the overflow path, and successive calls with
changing lengths and overflow counts check for stale state across graph
replays and context growth.  Batch tests compare each document's output
slice with an independent reference encoding, covering seams within digit
runs, contractions, and special-token fragments as well as batches that
contain empty documents.  The artifact's correctness suite records expected
skips for unsupported configurations as skips rather than converting them
into passes.

These results establish reference equivalence for the evaluated
configurations and encode modes, not for arbitrary tokenizer files or for
all possible inputs.  Chat templates, truncation, and padding remain outside
this bare-encoding comparison.  Separate Gemma tests specify fallback tokens
for malformed raw bytes.  They test a byte-input contract rather than
equivalence with a reference API that accepts Unicode text.

\takeaway{\sys reproduces the reference token IDs in every cell of the exactness
campaign, on both builds, and in every reported measurement.}

\subsection{How does \sys compare with CPU and GPU tokenizers?}
\label{sec:eval-gputok}
\label{sec:eval-standalone}

\smartparagraph{Prior GPU tokenizers.}
The natural first comparison is with the existing GPU tokenizer, so we built
GPUTOK~\cite{kadamba2026gputok} from source and verified it on the same
corpus before timing it, the same gate we apply to every system.  Two
properties disqualify it from the latency comparison.  First, on the
one-million-token books input it emits 989,225 tokens rather than 1,000,000:
it drops the pre-tokenizer that GPT-2's regular expression supplies, and it
first diverges at index 130, where it merges two newline tokens the
reference keeps separate.  Its bundled self-check does not catch this,
because it verifies the encoder against its own in-file greedy reference
rather than against the model's tokenizer, so it certifies internal
consistency of a different tokenization.  Second, its encode path cannot be
captured as a graph: it byte-encodes on the host, allocates buffers of
runtime-dependent size, issues synchronous copies, and synchronizes on an
event on every call.  We therefore compare \sys with GPUTOK on the property
this paper is about, exactness against the reference, where GPUTOK fails,
and leave its latency to its own report.

\smartparagraph{CPU baselines.}
Figure~\ref{fig:standalone-cpu-gpu} compares \sys with the CPU tokenizers
that produce the reference sequence on the Llama~3 vocabulary, namely
 Hugging Face tokenizers, tiktoken, gigatoken, and TokTier, each through its
single-document call.  TokTier runs in its default auto mode, which sends
inputs of at least \qty{64}{\kibi\byte} to its GPU kernel and smaller ones
to its \texttt{fast\_cpu} engine, with its shipped prebuilt kernel.  We
record the route of every timed call, and every input from the 32k file up
ran on TokTier's GPU, so its GPU curve is the shipped kernel, not a
fallback.  All curves except one are warm, re-encoding text the engine
has already seen, a setting that favours engines with caches.  The
exception is a second TokTier curve (dashed) on never-seen text, and
Figure~\ref{fig:cold-warm} quantifies the cold penalty for every engine.

Across four decades of input length, the figure shows two regimes.  Below
the crossover, \sys serves the input on its CPU route, whose engine is
gigatoken (open markers).  There it runs within 8\% of gigatoken on every
cell except the 128-token multilingual input (\qty{2.5}{\micro\second}
against \qty{2.2}{\micro\second}) and is ahead of warm and cold TokTier,
tiktoken, and Hugging Face at every size.  From the 32k input on books and
the 64k input on code and multilingual chat, the graph takes over and
\sys's lead grows with the input.  On the largest inputs, 0.63 to 0.91
million Llama~3 tokens, its complete call takes
 \qtyrange{1.70}{1.76}{\milli\second}.  This is $7.9$--$12.9\times$ faster
than TokTier's GPU route, $12.9$--$19.4\times$ faster than its
\texttt{fast\_cpu} engine (measured alone, not plotted),
$202$--$313\times$ faster than tiktoken, and $865$--$1295\times$ faster
than the Hugging Face tokenizers library, the reference implementation for
this vocabulary.  TokTier's GPU route beats its own CPU engine from the
64k input on but is slower than gigatoken at every size, and \sys is
$3.1$--$12.9\times$ faster than that route wherever it runs on the GPU.

Among the CPU tokenizers, gigatoken is the strongest baseline and merits a
careful comparison.  It matches the reference sequence on every
configuration it supports.  It is also the engine inside TokTier's
\texttt{fast\_cpu} backend, rebuilt and statically linked, so these two
baselines are one tokenizer measured through two wrappers, and the gap
between them reflects TokTier's routing and certificate machinery rather
than a different algorithm.  On the largest single document, \sys leads
gigatoken by $4.7$--$6.0\times$ across the three corpora
(Figure~\ref{fig:standalone-cpu-gpu}).

This comparison has two limits.  First, it compares latency between one GPU
and the host platform and is not normalized for cost or power.  Second,
latency is not throughput.  Given a large enough payload and all 96
hardware threads, gigatoken's aggregate batch throughput reaches
\qty{5.3}{\giga\byte\per\s}, above both \sys's transfer-bound
single-document complete call (\qty{2.2}{\giga\byte\per\s}) and its graph
alone (\qty{4.2}{\giga\byte\per\s}), but it occupies every hardware thread
on the host where \sys uses one.  That measurement is warm and its payload
repeats a \qty{13}{\mega\byte} corpus, so it is an upper bound on what a
cache-dependent engine achieves on repeated text, and we did not measure it
cold at that payload.  \sys targets single-stream latency and long
contexts rather than bulk offline throughput on a many-core host, and that
is where its advantage lies.

The graph-only series explains why short inputs go to the CPU route: on
the smallest input, a replay has a floor of
\qtyrange{0.06}{0.08}{\milli\second}, consisting of one launch, one
synchronization, and the fixed work of the graph's kernels, which a short
input cannot amortize.  By comparison, gigatoken encodes a 128-token input
in about \qty{2}{\micro\second}.  The other vocabularies follow the same
curve on books, with graph floors of \qtyrange{0.055}{0.083}{\milli\second}.
On the largest input (one million GPT-2 tokens and 0.91--0.92 million for
the others), the complete call takes from \qty{1.7}{\milli\second} on
Llama~3 to \qty{3.2}{\milli\second} on Qwen~2.5, whose NFC normalization
runs on the host.

The gap between the two \sys curves shows the cost of returning results to
the host.  For Llama~3 on the largest books input, graph replay takes about
\qty{1.0}{\milli\second} and the complete call about
\qty{1.8}{\milli\second}; the difference is transfer and output
materialization, which a device-resident consumer avoids.

\smartparagraph{Warm versus cold input.}
Figure~\ref{fig:cold-warm} measures the penalty on documents that no call
has submitted before.  Each timed call receives a new window of the same
length, drawn at random from \qty{256}{\mega\byte} of unique text of the
same kind, while every engine keeps its caches across calls as a server
would, so frequent words recur as they do in real traffic.  Every engine
receives its own string objects with their UTF-8 form already computed, as
on a warm call, so the ratio isolates tokenization.  The sweep covers 128
to 100k tokens; between its last two points no ratio moves by more than
$1.8\times$, and Figure~\ref{fig:standalone-cpu-gpu} carries absolute
latency to one million.

The CPU engines cache pre-tokens, so on never-repeated documents gigatoken
pays a median $4.5\times$, TokTier's \texttt{fast\_cpu} engine $2.6\times$,
and its auto mode $2.4\times$, the last near $1\times$ only where its GPU
kernel runs.  These CPU ratios are lower bounds: most cells of the CPU
sweeps re-read a region an earlier cell had seen, so the true first-sight
penalty is at least what is plotted.  \sys's graph pays only $1.03\times$
because it has no content cache to miss.  \sys as shipped pays a
CPU-engine penalty of the same order on the short inputs it routes to the
CPU and its graph's penalty everywhere else.  In the figure, open markers
are boundary cells where the router split the cold arm across both
routes, and the single dip below one is a warm-arm artifact, not a cold-arm
speedup.
Warm comparisons therefore understate \sys's advantage above the crossover,
and the crossover itself is measured on never-repeated text
(Section~\ref{sec:eval-cpu-route}).

\takeaway{Above the crossover, \sys's lead grows with input length and reaches
$4.7$--$6.0\times$ over gigatoken on the largest inputs.}

\subsection{What do the CPU route and the dispatcher add?}
\label{sec:eval-cpu-route}

\smartparagraph{Routing between two exact engines.}
A graph replay has a fixed cost that a short input cannot amortize, so \sys
serves short inputs on its CPU route (Section~\ref{sec:impl}) and routes
each input to the cheaper of the two engines.  We define the crossover as the largest size at
which the CPU route's median does not exceed the graph's, measured on
never-repeated windows drawn at random from \qty{256}{\mega\byte} each of
books, code, and multilingual Wikipedia.  It is \qty{16}{\kibi\byte} for
mostly-ASCII text and \qty{4}{\kibi\byte} when more than 1/64 of the bytes
start a non-ASCII character (typographic punctuation excluded), and it is
identical for the four byte-level vocabularies.  Between these limits and
eight times the ASCII limit, the router compares the costs it has recently
observed on each route for that content and size, because the CPU engine's
cost depends on how often it has seen the words while the graph's does not.
Below the crossover the CPU engine serves the call, so its latency is
that of the CPU route, and the complete-call series in
Figure~\ref{fig:standalone-cpu-gpu} coincides with gigatoken there.  The
router guarantees that the caller never pays more than the better of the two
exact engines, per input and in content-sensitive cases, without the caller
choosing.

\smartparagraph{Many concurrent requests.}
A serving frontend tokenizes many independent requests at once, the case
for which the library provides the dispatcher of Section~\ref{sec:impl}.
We run it with $k$ CPU engines, and its GPU queue uses no batching timer.
Its contexts wait for the GPU by sleeping rather than spinning and copy an
upper bound of the output with the replay, so a batch costs the host one
wake-up.  We offered an open-loop load of 1--16~KiB requests (log-uniform,
equal thirds of the three never-repeated corpora, Gamma-distributed
arrivals) to 64 request threads confined to 8~CPUs, for 3~s after 1~s of
warm-up per point.  We compare three policies, gigatoken on up to all 8
cores, the GPU queue alone, and the adaptive policy with $k=2$.

\begin{table}[t]
  \centering
  \small
  \caption{Concurrent 1--16~KiB requests on 8~CPUs: median latency and host
  CPU time per request.  The GPU queue saturates near 60k requests/s,
  gigatoken on 8 cores near 80k, the adaptive policy at 77k.}
  \label{tab:dispatch}
  \setlength{\tabcolsep}{4.5pt}
  \begin{tabular}{@{}rrrrrrr@{}}
    \toprule
    & \multicolumn{2}{c}{gigatoken, 8 cores} & \multicolumn{2}{c}{GPU queue}
    & \multicolumn{2}{c}{adaptive, $k=2$} \\
    \cmidrule(lr){2-3}\cmidrule(lr){4-5}\cmidrule(l){6-7}
    req/s & p50 & CPU & p50 & CPU & p50 & CPU \\
    \midrule
    10k & \qty{90}{\micro\second} & \qty{79}{\micro\second} & \qty{294}{\micro\second} & \qty{27}{\micro\second} & \qty{95}{\micro\second} & \qty{49}{\micro\second} \\
    20k & \qty{83}{\micro\second} & \qty{71}{\micro\second} & \qty{406}{\micro\second} & \qty{17}{\micro\second} & \qty{98}{\micro\second} & \qty{42}{\micro\second} \\
    40k & \qty{80}{\micro\second} & \qty{69}{\micro\second} & \qty{649}{\micro\second} & \qty{12}{\micro\second} & \qty{192}{\micro\second} & \qty{32}{\micro\second} \\
    60k & \qty{92}{\micro\second} & \qty{69}{\micro\second} & \qty{14.7}{\milli\second} & \qty{8}{\micro\second} & \qty{525}{\micro\second} & \qty{25}{\micro\second} \\
    \bottomrule
  \end{tabular}
\end{table}

Table~\ref{tab:dispatch} separates two resources.  Batching requests on the
GPU is how \sys gets its long-context speed, and at the same scale it costs
little host CPU: combining concurrent requests on the GPU cuts the host CPU
time per request by $2.9$--$8.2\times$ relative to gigatoken, because every
request in a replay shares its staging and launch, while a gigatoken core is
occupied for the full duration of each request it serves.  The price is in
the other resource: the single replay stream adds queueing, raising the
median latency from about \qty{0.09}{\milli\second} to
\qtyrange{0.3}{0.65}{\milli\second} and saturating near 60k requests/s,
below gigatoken's 80k on eight cores.  The adaptive policy exists for
precisely this reason: it takes requests to the GPU only where the replay is
not the bottleneck, so at 40k requests/s it spends $2.2\times$ less CPU per
request than gigatoken while keeping a median of
\qty{0.19}{\milli\second}, and it sustains 77k requests/s.  Which resource
is scarce is a property of the deployment: a frontend whose CPUs are also
serving agents and retrieval buys the CPU saving at a latency cost it can
absorb, while a frontend that tokenizes nothing else should pin short
requests to the CPU route alone.

\takeaway{Below the crossover, \sys matches gigatoken on latency by choosing,
per input, between two exact engines, and it buys that guarantee with
content-adaptive routing the caller does not have to manage. For short
requests the dispatcher trades a little latency for $2.9$--$8.2\times$ less
host CPU per request, a trade that is worthwhile where the frontend's CPUs
serve agents and other work alongside tokenization. On long prompts, the
graph is faster outright (Section~\ref{sec:eval-fleet}).}

\subsection{How does throughput scale with batch size?}
\label{sec:eval-batch}

Batching matters when a frontend already has several texts waiting, since
a single call can then amortize \sys's fixed invocation cost and expose more
independent work to the GPU.  Figure~\ref{fig:batch-scaling} measures this
effect for 8k-token Llama~3 documents on all three corpora, comparing each
implementation through its native batch interface (and a batch of one
through its single-document call).  TokTier's auto mode routes each document by its own size, so
every 8k-token document stays on its CPU engine, and \texttt{fast\_cpu}
measured directly agrees with auto mode within 8\%, so one curve stands for
both modes.  All complete-call outputs are checked per document against the
reference before timing.  The graph-only series replays the same captured
pipeline after untimed batch staging, and its outputs are checked by an
untimed token-and-offset readback from the same staged benchmark context.

\sys's complete-call throughput rises from 8,000--16,600 documents/s at
batch~1 to 41,000--66,000 documents/s at batch~256, and the amortized time
per document falls from \qtyrange{0.060}{0.124}{\milli\second} to
\qtyrange{0.015}{0.025}{\milli\second}.  At batch~1, a single 8k-token
document is a single-document call, and the router chooses between the CPU
route and a graph that one document leaves mostly idle.  gigatoken is
faster in this case, with \sys running at $0.41$--$0.93\times$ its speed
and slowest on multilingual chat, where the router keeps the graph.  From
batch~8 on, \sys is the fastest implementation on every corpus.  Its lead
over gigatoken peaks at $2.9$--$3.8\times$ at batch~32--64 and narrows to
$1.1$--$1.4\times$ at batch~256, where gigatoken's parallel batch path
fills the host's cores.  At batch~256, \sys leads TokTier by
$10$--$13\times$ in either mode, tiktoken by $36$--$47\times$, and Hugging
Face by $45$--$59\times$.

Throughput levels off beyond batch~64, changing by between $-20\%$ (books,
whose 256-document batch fills an 8~MiB graph) and $+11\%$ from 64 to 256.
Larger batches therefore offer little benefit and, in an online system,
would mainly add queueing delay while the batch fills; the benchmark here
assumes the batch is already available and does not include that delay.

The CPU engines flatten earlier.  They saturate their thread pools by
batch~8--64, and TokTier declines beyond that point, whereas one graph
replay covers the whole batch.  One property of the benchmark favours the
CPU engines.  Each batch repeats one document, which gives them
pre-token-cache hits worth 15--22\% on every document after the first, a
benefit that \sys's device path does not receive.

\takeaway{From batch~8 on, \sys is the fastest implementation on every corpus,
and its lead over gigatoken peaks at batch~32--64.}

\subsection{How does \sys compare with TokTier on TokTier's own benchmark?}
\label{sec:eval-toktier}

TokTier~\cite{zhang2026toktier}, developed concurrently and independently
of our work, addresses the same problem of exact tokenization with a GPU
route.  We reran its single-request experiment on our host with its
released package~0.2.9, following its benchmark protocol.  Contexts are
rendered from public SWE-smith \cite{yang2025swesmith} coding-agent
trajectories, each timed call encodes a context that no engine has seen
before, each engine runs in its own process on one core with CPU engines
single-threaded, and every output is checked against Hugging Face
tokenizers.  As in TokTier's table, gigatoken gets a new object, and hence
an empty cache, for every call, with construction excluded.  Because
returning a Python list costs a caller more than returning an array, we
compare only engines that deliver the same kind of output:
Figure~\ref{fig:single-request} shows the list-returning calls, and
gigatoken, whose native call returns a NumPy array, is converted to a list
as vLLM's adapter does.  Our headline comparisons time every system as
released, and we note the paper's number where the two differ.  TokTier's
paper reports \qty{0.38}{\milli\second} at one million characters for a
CUDA-graph path with array delivery that its released package does not
contain; on Llama~3.1, the released GPU route takes
\qty{4.9}{\milli\second} with list delivery on our host, and \sys's array
call takes \qty{1.35}{\milli\second}.

\smartparagraph{Single requests.}
Figure~\ref{fig:single-request} reproduces TokTier's single-request table
from 100K to 4.4 million characters.  At the median, \sys is
$2.0$--$2.4\times$ faster than TokTier's GPU route on Llama~3.1 and
$2.0$--$18.1\times$ faster on Qwen, where that route takes
\qty{33.7}{\milli\second} at one million characters.  It is also
$2.7$--$3.7\times$ and $2.6$--$3.9\times$ faster than gigatoken, and the gap
widens as the context grows.  From one million characters up, \sys also has
the lowest P99 of every engine: \qty{14.2}{} and \qty{14.4}{\milli\second}
at 4.4 million, against TokTier's \qty{27.6}{\milli\second} on Llama~3.1 and
gigatoken's \qty{51.0}{\milli\second} on Qwen.

This tail depends on merging long pre-tokens in batches.  One of these
contexts holds a 10,000-character run of a single character, a single
pre-token whose merges the one-merge-per-step path would apply one at a
time.  The final build's merge policy applies all lowest-rank merges of a
round together, with a check that keeps the result bit-identical to
sequential merging.  With it, that context takes the median time; the final
build runs every result in this section, including the fleet experiment of
Section~\ref{sec:eval-fleet}.  At 100K characters, \sys as shipped has a P99
of \qty{0.91}{\milli\second} on Llama~3.1, below gigatoken's \qty{1.21}{}
but just above TokTier's \qty{0.83}{\milli\second}.  With every input on
the GPU and a blocking host wait instead of a spinning one, the median of
that call is \qty{0.44}{\milli\second} and its P99 \qty{0.63}{\milli\second}.

\takeaway{On TokTier's own benchmark, \sys has the lowest median latency of every
engine at every size in both families, and the lowest P99 from one million
characters up.}

\subsection{Does the speedup reach the serving critical path?}
\label{sec:eval-integration}

\providecommand\servnum[1]{\ifcsname servnum/#1\endcsname\csname servnum/#1\endcsname\else\textbf{??#1}\GenericWarning{}{Undefined serving number #1}\fi}
\expandafter\def\csname servnum/samples\endcsname{5}
\expandafter\def\csname servnum/samples-pooled\endcsname{35}
\expandafter\def\csname servnum/speedup99-32k\endcsname{2.43}
\expandafter\def\csname servnum/tokonly99-32k\endcsname{1.88}
\expandafter\def\csname servnum/delivery99-32k\endcsname{16}
\expandafter\def\csname servnum/prefill99-gpu-32k\endcsname{37}
\expandafter\def\csname servnum/ttft99-cpu-32k\endcsname{126}
\expandafter\def\csname servnum/tok-cpu-32k\endcsname{59.1}
\expandafter\def\csname servnum/tokmax-cpu-32k\endcsname{64.7}
\expandafter\def\csname servnum/tokfirst-cpu-32k\endcsname{62.6}
\expandafter\def\csname servnum/ttft99-tiktoken-32k\endcsname{69.5}
\expandafter\def\csname servnum/tok-tiktoken-32k\endcsname{20.8}
\expandafter\def\csname servnum/tokmax-tiktoken-32k\endcsname{26.1}
\expandafter\def\csname servnum/tokfirst-tiktoken-32k\endcsname{26.1}
\expandafter\def\csname servnum/ttft99-gigatoken-32k\endcsname{53.2}
\expandafter\def\csname servnum/tok-gigatoken-32k\endcsname{1.6}
\expandafter\def\csname servnum/tokmax-gigatoken-32k\endcsname{3.0}
\expandafter\def\csname servnum/tokfirst-gigatoken-32k\endcsname{3.0}
\expandafter\def\csname servnum/ttft99-toktier_auto-32k\endcsname{54.0}
\expandafter\def\csname servnum/tok-toktier_auto-32k\endcsname{1.4}
\expandafter\def\csname servnum/tokmax-toktier_auto-32k\endcsname{139}
\expandafter\def\csname servnum/tokfirst-toktier_auto-32k\endcsname{139}
\expandafter\def\csname servnum/ttft99-gpu-32k\endcsname{51.9}
\expandafter\def\csname servnum/tok-gpu-32k\endcsname{2.9}
\expandafter\def\csname servnum/tokmax-gpu-32k\endcsname{12.7}
\expandafter\def\csname servnum/tokfirst-gpu-32k\endcsname{12.7}
\expandafter\def\csname servnum/speedup99-64k\endcsname{2.35}
\expandafter\def\csname servnum/tokonly99-64k\endcsname{1.91}
\expandafter\def\csname servnum/delivery99-64k\endcsname{29}
\expandafter\def\csname servnum/prefill99-gpu-64k\endcsname{104}
\expandafter\def\csname servnum/ttft99-cpu-64k\endcsname{303}
\expandafter\def\csname servnum/tok-cpu-64k\endcsname{138}
\expandafter\def\csname servnum/tokmax-cpu-64k\endcsname{156}
\expandafter\def\csname servnum/tokfirst-cpu-64k\endcsname{123}
\expandafter\def\csname servnum/ttft99-tiktoken-64k\endcsname{166}
\expandafter\def\csname servnum/tok-tiktoken-64k\endcsname{41.8}
\expandafter\def\csname servnum/tokmax-tiktoken-64k\endcsname{45.8}
\expandafter\def\csname servnum/tokfirst-tiktoken-64k\endcsname{41.9}
\expandafter\def\csname servnum/ttft99-gigatoken-64k\endcsname{129}
\expandafter\def\csname servnum/tok-gigatoken-64k\endcsname{2.7}
\expandafter\def\csname servnum/tokmax-gigatoken-64k\endcsname{3.8}
\expandafter\def\csname servnum/tokfirst-gigatoken-64k\endcsname{3.4}
\expandafter\def\csname servnum/ttft99-toktier_auto-64k\endcsname{129}
\expandafter\def\csname servnum/tok-toktier_auto-64k\endcsname{2.2}
\expandafter\def\csname servnum/tokmax-toktier_auto-64k\endcsname{4.8}
\expandafter\def\csname servnum/tokfirst-toktier_auto-64k\endcsname{4.8}
\expandafter\def\csname servnum/ttft99-gpu-64k\endcsname{129}
\expandafter\def\csname servnum/tok-gpu-64k\endcsname{3.8}
\expandafter\def\csname servnum/tokmax-gpu-64k\endcsname{15.2}
\expandafter\def\csname servnum/tokfirst-gpu-64k\endcsname{15.2}
\expandafter\def\csname servnum/speedup99-128k\endcsname{2.06}
\expandafter\def\csname servnum/tokonly99-128k\endcsname{1.76}
\expandafter\def\csname servnum/delivery99-128k\endcsname{56}
\expandafter\def\csname servnum/prefill99-gpu-128k\endcsname{283}
\expandafter\def\csname servnum/ttft99-cpu-128k\endcsname{671}
\expandafter\def\csname servnum/tok-cpu-128k\endcsname{284}
\expandafter\def\csname servnum/tokmax-cpu-128k\endcsname{337}
\expandafter\def\csname servnum/tokfirst-cpu-128k\endcsname{294}
\expandafter\def\csname servnum/ttft99-tiktoken-128k\endcsname{406}
\expandafter\def\csname servnum/tok-tiktoken-128k\endcsname{84.0}
\expandafter\def\csname servnum/tokmax-tiktoken-128k\endcsname{97.1}
\expandafter\def\csname servnum/tokfirst-tiktoken-128k\endcsname{84.0}
\expandafter\def\csname servnum/ttft99-gigatoken-128k\endcsname{326}
\expandafter\def\csname servnum/tok-gigatoken-128k\endcsname{4.8}
\expandafter\def\csname servnum/tokmax-gigatoken-128k\endcsname{10.7}
\expandafter\def\csname servnum/tokfirst-gigatoken-128k\endcsname{5.9}
\expandafter\def\csname servnum/ttft99-toktier_auto-128k\endcsname{329}
\expandafter\def\csname servnum/tok-toktier_auto-128k\endcsname{3.7}
\expandafter\def\csname servnum/tokmax-toktier_auto-128k\endcsname{8.3}
\expandafter\def\csname servnum/tokfirst-toktier_auto-128k\endcsname{4.0}
\expandafter\def\csname servnum/ttft99-gpu-128k\endcsname{326}
\expandafter\def\csname servnum/tok-gpu-128k\endcsname{5.8}
\expandafter\def\csname servnum/tokmax-gpu-128k\endcsname{31.8}
\expandafter\def\csname servnum/tokfirst-gpu-128k\endcsname{31.8}
\expandafter\def\csname servnum/speedup0-range\endcsname{1.02--1.04}
\expandafter\def\csname servnum/speedup0-pct\endcsname{2--4}
\expandafter\def\csname servnum/speedup99-range\endcsname{2.06--2.43}
\expandafter\def\csname servnum/tokonly99-range\endcsname{1.76--1.91}
\expandafter\def\csname servnum/share0-range\endcsname{1.8--3.3}
\expandafter\def\csname servnum/share99-range\endcsname{42--47}
\expandafter\def\csname servnum/ttftgap99-fast\endcsname{4}
\expandafter\def\csname servnum/ttftgap99-cpu\endcsname{143}
\expandafter\def\csname servnum/tokratio-cpu\endcsname{20.2--49.0}
\expandafter\def\csname servnum/tokinvratio-cpu\endcsname{0.0}
\expandafter\def\csname servnum/tokgap-cpu\endcsname{-277.8---56.1}
\expandafter\def\csname servnum/ttftgap99-tiktoken\endcsname{34}
\expandafter\def\csname servnum/tokratio-tiktoken\endcsname{7.1--14.5}
\expandafter\def\csname servnum/tokinvratio-tiktoken\endcsname{0.1}
\expandafter\def\csname servnum/tokgap-tiktoken\endcsname{-78.2---17.9}
\expandafter\def\csname servnum/ttftgap99-gigatoken\endcsname{3}
\expandafter\def\csname servnum/tokratio-gigatoken\endcsname{0.6--0.8}
\expandafter\def\csname servnum/tokinvratio-gigatoken\endcsname{1.2--1.8}
\expandafter\def\csname servnum/tokgap-gigatoken\endcsname{1.0--1.3}
\expandafter\def\csname servnum/ttftgap99-toktier_auto\endcsname{4}
\expandafter\def\csname servnum/tokratio-toktier_auto\endcsname{0.5--0.6}
\expandafter\def\csname servnum/tokinvratio-toktier_auto\endcsname{1.6--2.2}
\expandafter\def\csname servnum/tokgap-toktier_auto\endcsname{1.6--2.1}

We connect \sys to vLLM~0.27.0 through its tokenizer factory without
modifying vLLM source and serve 32k--128k-token prompts one at a time.  With
a cold prefix cache, replacing vLLM's stock tokenizer lowers mean
\gls{ttft} by \servnum{speedup0-pct}\%, the small share tokenization holds
while prefill dominates.  At 99\% prefix reuse prefill is removed and the
frontend is exposed.  The tokenization-only gain is
\qtyrange{1.76}{1.91}{\times} at all three prompt sizes, and the mean
\gls{ttft} gain is larger, \servnum{speedup99-range}$\times$.  The
difference comes from vLLM's incremental detokenizer, which decodes the
whole prompt on the first decode step when the tokenizer is a
\texttt{PreTrainedTokenizerFast}, as the stock tokenizers are and the \sys
proxy is not.  This adds \qty{16}{\milli\second}, \qty{29}{\milli\second},
and \qty{56}{\milli\second} to the stock arm's first token at 32k, 64k, and
128k tokens, and we attribute it to the stock arm rather than to \sys.  We
pool five requests per cell, the most that keep the shared prefix cached,
and report means without confidence intervals.  Here gigatoken and TokTier
come within \servnum{ttftgap99-fast}\% of \sys.  The next subsection
measures the same mechanism with 10--40 requests per length and confirms
that in one engine the binding cost is the stock frontend rather than the
choice of fast tokenizer, with median gaps of 3--18\%.

\takeaway{At 99\% prefix reuse, replacing vLLM's stock tokenizer with \sys lowers mean
\gls{ttft} by \servnum{speedup99-range}$\times$ for 32k--128k-token prompts.}

\subsection{How much of vLLM's time to first token is tokenization?}
\label{sec:eval-burst}

The vLLM experiment above isolates the mechanism.  This experiment instead
measures how much of a long-context request's \gls{ttft} each exact
tokenizer in this evaluation takes inside the same serving engine, from 32k
to one million tokens.  On one H100 NVL with prefix caching enabled,
vLLM~0.27.0 serves three models.  Llama-3.1-8B-Instruct (bf16) serves prompts of 32k,
64k, and 131k tokens.  Qwen3-32B, the model of the frontend experiment in
Section~\ref{sec:eval-fleet}, runs with FP8 weights and KV cache and with
YaRN extending its context to 262k tokens, and serves prompts of 32k to
262k tokens.  Qwen2.5-7B-Instruct-1M, whose tokenizer is Qwen3's, serves
512k and one million tokens.  The servers are pinned to the CPU socket the
GPU is attached to, and the Qwen3-32B runs use 128-token KV blocks.  Only
the tokenizer changes between arms, which are stock Hugging Face, tiktoken,
TokTier in auto mode with its GPU kernel prebuilt, gigatoken, and \sys in
auto mode.  Each tokenizer is installed through the same tokenizer-factory
hook and, before any timing, must return Hugging Face's token IDs for every
measured prompt in both single and batched calls.

\begin{figure*}[t]
  \centering
  \includegraphics[width=\textwidth]{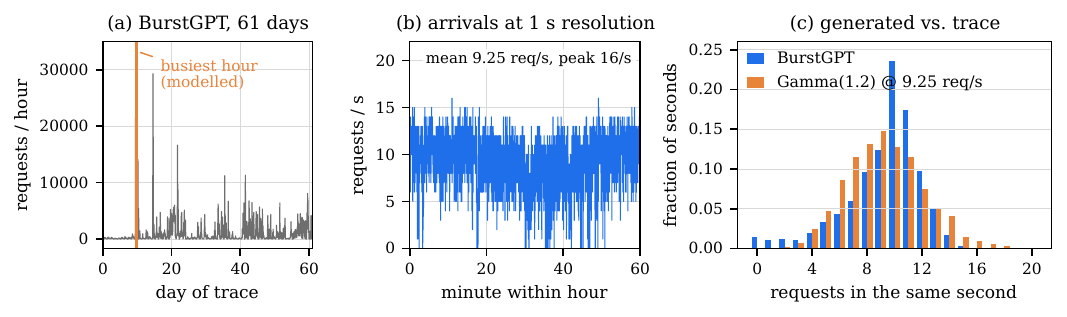}
  \caption{Arrival process derived from BurstGPT: (a)~the trace with its
  busiest hour marked, (b)~that hour per second, and (c)~concurrent arrivals
  of the fitted generator against the trace.}
  \label{fig:burstgpt}
  \Description{Three panels: a 61-day request-rate timeline with one hour
  highlighted, that hour's per-second arrivals, and a histogram comparing
  the generated arrival process to the trace.}
\end{figure*}

\smartparagraph{Workload.}
We size the burst from BurstGPT~\cite{burstgpt}, a 61-day production trace
of 1.4~million ChatGPT and GPT-4 requests, by fitting its busiest hour
(Figure~\ref{fig:burstgpt}).  The fit yields 9.25~requests/s with
Gamma-distributed inter-arrival gaps of shape 1.2, whose per-second arrival
counts have a coefficient of variation of 0.84, against 0.83 in the trace.
The generator thus matches the trace's rate and per-second dispersion.  Its
tail of simultaneous arrivals is slightly softer (44\% of generated seconds
carry at least 10 requests, against 58\% in the trace).  A softer tail is,
if anything, easier for the frontend to absorb, so the sustained loads
measured under it in Section~\ref{sec:eval-fleet} are conservative.

Here we measure each length one request at a time, which puts the tokenizer
on the critical path (40 requests per length on Qwen3-32B and 10 per length on
the other models).  The prompts follow the agentic pattern that motivates this
work, a long shared context followed by a short per-request turn such as a
tool result or a user message.  Each length has one shared prefix of
never-seen book text followed by a short turn (32--144 tokens), so
99.85--99.99\% of the prompt is cached.  We load the prefix into vLLM's
prefix cache by sending its token IDs, so no tokenizer sees any of the text
before its measured call, and every request adds a unique suffix.
gigatoken runs cold, as a fresh instance per call built at server start,
and TokTier runs with its document cache disabled, both checked on every
call.  \sys captures all of its graph size classes at server start.  Each
arm runs one pass per length without a warm-up call and generates 8 tokens
per request.

\begin{figure*}[t]
  \centering
  \includegraphics[width=\textwidth]{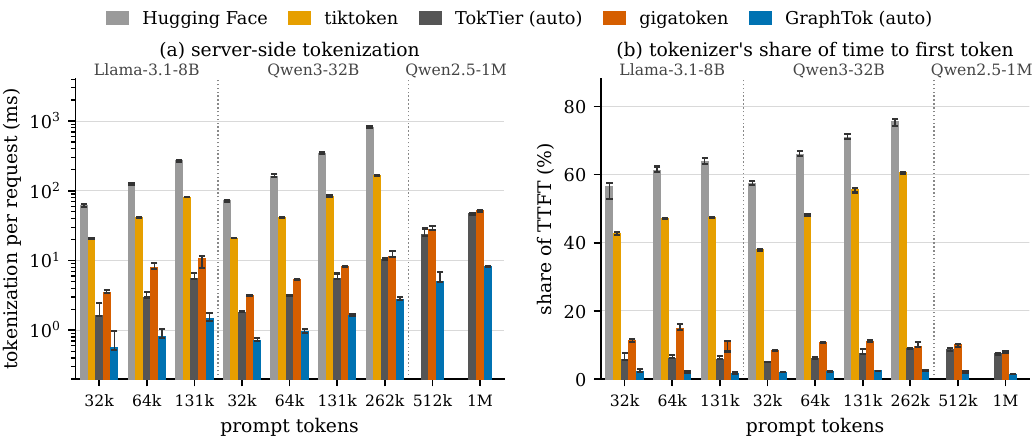}
   \caption{Tokenization inside vLLM, one long-context request at a time,
   never-seen text, 99.85--99.99\% of each prompt cached.  The left group is
   Llama-3.1-8B at 32k, 64k, and 131k tokens, the middle group Qwen3-32B
   (FP8) at 32k to 262k, and the right group Qwen2.5-7B-Instruct-1M at 512k
   and one million tokens.  (a)~The server-side tokenizer call per request
   (log scale).  (b)~That call's share of the request's \gls{ttft}.  Bars
   are medians and whiskers the 10th and 90th percentiles over 40 requests
   per length on Qwen3-32B and 10 on the others.  The sweep runs one pass
   per length without a warm-up call, so the bars are first-sight latencies;
   every arm sees the same condition.  Hugging Face and tiktoken were not
   run on the 1M-token model, where they take several seconds per request.}
  \label{fig:burst-ttft}
   \Description{Left: tokenization time per request for Llama-3.1-8B at 32k,
    64k and 131k tokens, Qwen3-32B at 32k to 262k, and Qwen2.5-7B-1M at 512k
    and 1M; GraphTok is lowest at every length.  Right: that call's share of
    time to first token, largest for the stock tokenizer and smallest for
    GraphTok.}
\end{figure*}

\smartparagraph{Result.}
\sys is the fastest tokenizer inside vLLM at every length
(Figure~\ref{fig:burst-ttft}(a)).  It tokenizes a request in a median of
\qtyrange{0.6}{2.8}{\milli\second} up to 262k tokens and in \qty{4.9}{} and
\qty{8.2}{\milli\second} at 512k and one million tokens.  This is
$2.5$--$5.8\times$ less than TokTier, the next fastest, $4.0$--$9.7\times$
less than gigatoken, $29$--$60\times$ less than tiktoken, and
$99$--$292\times$ less than Hugging Face.  With most of each prompt cached,
prefill no longer hides the tokenizer.  The stock tokenizer takes 56--76\%
of \gls{ttft} (Figure~\ref{fig:burst-ttft}(b)), the largest host-side share
reported for long-context vLLM serving~\cite{chung2026cpuslowdowns};
tiktoken takes 38--60\%, gigatoken up to 15\%, and TokTier up to 9\%,
whereas \sys takes 1.4--2.5\%.

As a result, \sys has the lowest median \gls{ttft} at every length.  Its
median is $3.5$--$9.8\times$ below the stock tokenizer's, with the gap
growing with length (\qty{110}{} against \qty{1077}{\milli\second} at 262k
tokens), $1.5$--$2.5\times$ below tiktoken's, 3--13\% below TokTier's, and
3--18\% below gigatoken's; its 90th percentile is also the lowest at every
length.  These gaps depend on how little other per-request work the engine
performs.  With vLLM's default 16-token KV blocks, prefill and per-request
block bookkeeping grow, so at 131k tokens on Llama-3.1-8B (a 144-token turn)
\sys's lead shrinks to $4.7\times$ over the stock tokenizer and 10\% over
gigatoken.

Under load, a single engine saturates before any fast tokenizer does.  With
Poisson arrivals of 10 to 50 requests/s at 131k tokens on Llama-3.1-8B, one
H100 completes about 9 requests/s, and the median \gls{ttft} of \sys and of
gigatoken (with its cache reset before every call) stay within 15\% of each
other at every rate, in either direction.

\takeaway{Inside vLLM, \sys takes only 1.4--2.5\% of \gls{ttft}, against
56--76\% for the stock tokenizer. A tokenizer that costs a few milliseconds
is thus off the critical path of one engine on one GPU. Its cost becomes a
capacity limit where one frontend tokenizes for many GPUs, which the next
experiment measures.}

\subsection{How many GPUs can one frontend feed?}
\label{sec:eval-fleet}

Systems that serve a GPU fleet place a shared frontend in front of it.
NVIDIA Dynamo's frontend renders the chat template and tokenizes each
prompt, and only then can its KV-aware router hash the token blocks and
select a worker~\cite{nvidia2026dynamokvrouter}.  Every request for every
GPU therefore passes through one tokenizer, and prefix caching, which lets
each GPU turn requests around quickly, also raises the rate at which the
fleet sends requests into that tokenizer.  This experiment measures two
properties of that shared stage, the per-request tokenization cost under
load and the offered load a single frontend can sustain within a
\gls{ttft} target.

\smartparagraph{Setup.}
We run Dynamo~1.3.0 (commit \texttt{bae0051}) with its own frontend,
including its HTTP server, chat templating, KV router, and request plane,
and change only the tokenizer.  Dynamo's default configuration encodes with
Hugging Face \texttt{tokenizers} and leaves its prefix tokenizer cache off.
\sys runs through the native backend of Section~\ref{sec:impl} in its auto
mode, in which the library's load-adaptive dispatcher combines concurrent
requests into GPU batches and would run short inputs on a CPU engine; every
request here is long enough for the GPU.  gigatoken, the fastest CPU
tokenizer of Section~\ref{sec:eval-standalone}, is a Rust library, and we
load it through the same native backend so that it runs inside the frontend
process exactly as \sys does, with its concurrency tuned for the frontend
and its cross-request pre-token cache emptied back to its vocabulary seed
before every request, a choice we return to below.
TokTier's GPU path exists only in its Python package, so TokTier runs in a
separate tokenizer process on the frontend's CPUs that the frontend reaches over a Unix socket.
Before any timing, every tokenizer must reproduce Hugging Face's token IDs
on the experiment's requests.

TokTier's package forwards any input containing an added token to its
reference backend, and every templated request contains
\texttt{<|im\_start|>}; as shipped, a 121k-token request therefore spends
\qty{227}{\milli\second} in the reference backend.  All arms are installed
through the same native-backend hook, which registers a tokenizer as an
exact engine for the vocabulary.  For TokTier, the hook splits the text at
the added-token boundaries, as Hugging Face itself does, encodes the text
between them with TokTier, and re-emits the boundary tokens, returning a
sequence that matches Hugging Face on every request.  This is the
documented integration path for tokenizers whose public entry point does
not accept added-token text, and it reduces the 121k-token request to
\qty{19.5}{\milli\second}.

The GPUs behind the frontend are simulated by Dynamo's mock engine, which
registers workers, publishes KV-cache events, and models scheduling and
prefix caching.  We run its timing model $1000\times$ faster than its
default so that no simulated GPU saturates and the frontend, the component
this paper changes, is the bottleneck under test, while the real H100 NVL
performs the tokenization being measured.  Each simulated GPU is offered
9.25 requests/s, the busiest-hour rate of BurstGPT, with the
Gamma-distributed arrivals of Section~\ref{sec:eval-burst}.  A fifth of the
requests continue one of the fleet's agent sessions, one per GPU, each with
a context of 121k, 512k, or one million tokens that the router keeps on the
worker caching it, followed by a new 2k-token turn.  The remaining requests
open a new session.  Contexts come from 2\,GB each of books (PG-19), source
code, and Wikipedia text in 16 languages.  A new session's context is text
that no earlier request in the run carried, and a continuing session
resends its context, as an agent resends its history.

A load generator written in Rust measures \gls{ttft} from each request's
scheduled arrival, so that queueing inside the frontend appears as latency,
and runs for 20~s per point.  The mock engine and the load generator run on
the host's other socket, and \sys runs on one H100 NVL.  We report load as
the request rate offered to the frontend, and also as the number of
simulated GPUs whose traffic that rate feeds.  The first is the measured
quantity; the second is a unit of offered load rather
than a measured prefill capacity.  A load is \emph{sustained} when at least
99\% of its requests complete within a P99 \gls{ttft} target.  Every
tokenizer is swept over the same offered loads, from the lowest up to the
first at which it saturates (P99 above \qty{10}{\second} or more than 10\%
of requests failed).  Loads with failures are marked separately, because
the percentile of the surviving requests would understate what the rest
experienced.

\begin{figure}[t]
  \centering
  \includegraphics[width=\columnwidth]{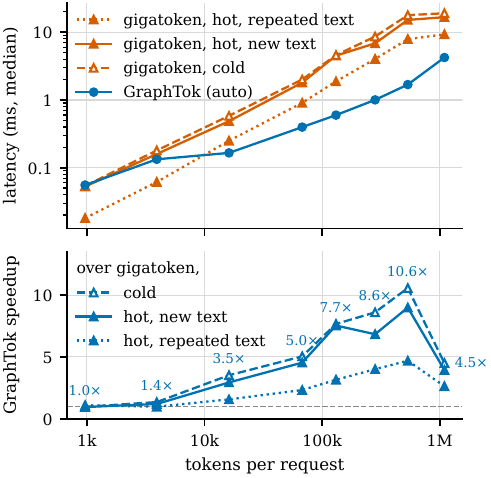}
  \caption{gigatoken with a hot and a cold pretoken cache, one request at a
  time on the Dynamo frontend (Qwen3 tokenizer, median of 40 documents
  per size, rotating books, code, and multilingual text, log--log).  Hot keeps
  the cache across requests, and cold empties it before every request.  \sys
  runs the same documents in its default auto mode, on one H100 NVL and on
  its CPU route below \qty{16}{\kibi\byte}.  Bottom:
  \sys's speedup (ratio of medians) over each gigatoken line, labelled
  against cold gigatoken.  All engines return identical token IDs on every
  request.}
  \label{fig:giga-hot-cold}
   \Description{A log--log plot of median latency against tokens per request
    from 1 thousand to 1 million.  On new text the cold and hot gigatoken lines
    are close, while hot gigatoken on repeated text is lower still; GraphTok is
    the lowest line from 16 thousand tokens up.  A second panel plots GraphTok's
    speedup over each gigatoken line.}
\end{figure}

\smartparagraph{gigatoken's cache.}
A continuing session resends text that gigatoken's cache has already
encoded, so as shipped it performs part of that work as cache lookups rather
than tokenization.  We empty the cache before every request to measure the
cost of tokenizing each request from scratch; Figure~\ref{fig:giga-hot-cold}
shows how much the cache would otherwise save.  \sys needs no cross-request
cache for its speed.  From 16k tokens up, it is $2.9$--$9.0\times$ faster
than gigatoken on new text and $1.6$--$4.7\times$ faster even on the text
that gigatoken's cache has seen.

\begin{figure*}[t]
  \centering
  \includegraphics[width=\textwidth]{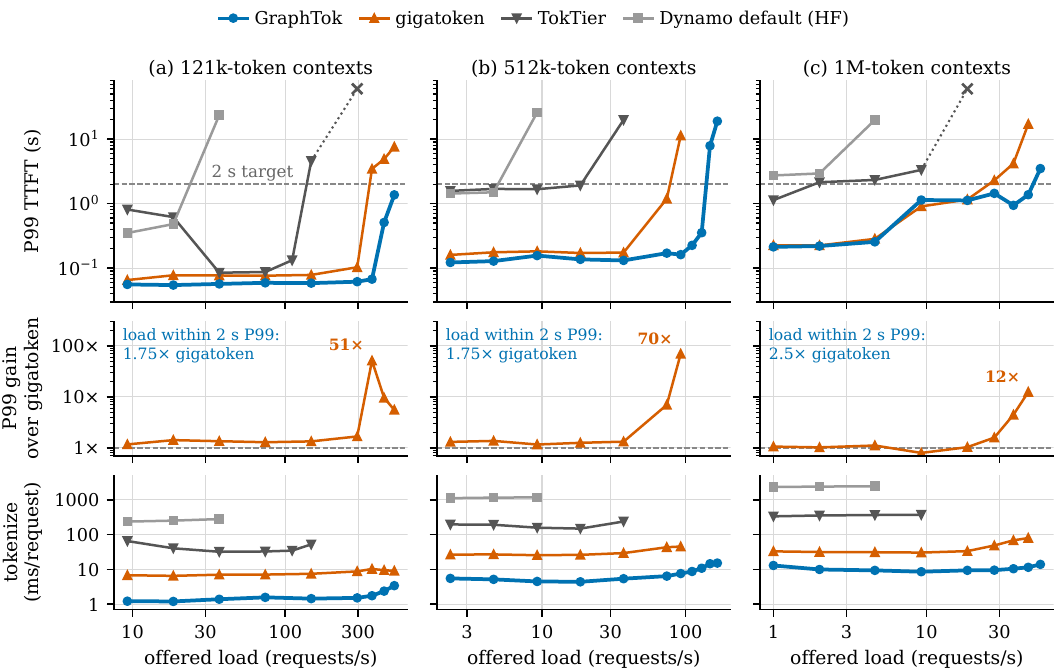}
  \caption{P99 \gls{ttft} of one Dynamo frontend against offered load
   (load expressed in GPUs at 9.25 requests/s each).  Every tokenizer is
   measured at the same
   loads, up to the first at which it saturates.  The dashed line is the
   \qty{2}{\second} P99 target; gigatoken runs with its pretoken cache emptied
   before every request (Figure~\ref{fig:giga-hot-cold}).  Middle: \sys's P99
   gain, gigatoken's P99 divided by \sys's at each load both completed, with
   the largest gain labelled together with the factor by which \sys raises the
   load sustained within the \qty{2}{\second} target.  Bottom: Dynamo's
   tokenize-stage time per request at every completed load.}
  \label{fig:dynamo-fleet}
   \Description{Three log--log panels plot, against offered load, P99 time to
    first token for 121k-, 512k- and 1M-token contexts (top), the ratio of
    gigatoken's P99 to GraphTok's (middle), and the per-request tokenize stage
    (bottom).  GraphTok is the lowest curve and the last to rise past the
    2-second target; the CPU tokenizers saturate at progressively lower loads.}
\end{figure*}

\smartparagraph{\sys sustains the most load at every length.}
Within the \qty{2}{\second} P99 target, the frontend sustains 518, 129.5,
and 46.25 requests/s with \sys at 121k, 512k, and one million tokens.  With
gigatoken it sustains 296, 74, and 18.5 ($1.75\times$, $1.75\times$, and
$2.5\times$ less), with TokTier 111, 18.5, and 1 ($4.7\times$, $7\times$, and
$46\times$ less), and with Dynamo's default tokenizer 18.5, 4.6, and none
($28\times$ less at 121k and 512k tokens, and at one million tokens even one
request per second waits \qty{2.7}{\second} at the P99).  Below these loads,
\sys's P99 is also the lowest.  At 121k tokens it stays at \qtyrange{55}{68}{\milli\second}
up to 370 requests/s, while gigatoken's is
\qtyrange{67}{172}{\milli\second} at the loads below 370.  Over the loads at
which TokTier and the default still meet the \qty{2}{\second} target, their
P99s are \qtyrange{85}{810}{} and \qtyrange{350}{482}{\milli\second}.

\smartparagraph{\sys holds the tail where the others break.}
Each tokenizer's P99 stays flat until the frontend can no longer keep up
with the offered load and then rises by orders of magnitude within one or
two load steps.  \sys reaches this knee last.  gigatoken already
misses the target at 370 requests/s and 121k tokens, with a P99 of
\qty{3.5}{\second} against \sys's \qty{68}{\milli\second} ($51\times$).  At
148 requests/s, TokTier's P99 is \qty{4.6}{\second}, $78\times$ \sys's, and
at 37 requests/s the default's is \qty{23.5}{\second}, $12\times$ its own
\qty{2}{\second} target and the largest single miss at 121k tokens.  At
512k tokens and 111 requests/s, gigatoken's P99 is \qty{14.4}{\second}
against \sys's \qty{226}{\milli\second}, $64\times$ lower.  At one million
tokens and 46.25 requests/s, gigatoken's P99 is \qty{17.1}{\second} against
\sys's \qty{1.4}{\second}.  These ratios
are taken where one tokenizer has passed its knee and the other has not, so
their magnitude depends on the load grid and the \qty{20}{\second} window of
each point; the middle row of Figure~\ref{fig:dynamo-fleet} labels the
largest such ratio at each length.  The sustained-load ratios above are the
primary result, and the tail ratios show how sharply the systems separate
past the knee.

The difference comes from tokenization under load.  Over every load it was
offered, Dynamo's tokenize-stage timer reads \qtyrange{1.2}{4.1}{},
\qtyrange{4.4}{15.2}{}, and \qtyrange{8.6}{15.8}{\milli\second} per request
for \sys at the three lengths, about $200\times$ less than the default.
gigatoken's time grows with load, from \qty{22}{} to
\qty{38}{\milli\second} at 512k tokens and from \qty{27}{} to
\qty{73}{\milli\second} at one million, and TokTier's is
\qtyrange{32}{65}{\milli\second} at 121k tokens.

\smartparagraph{Dynamo, not \sys, sets the limit.}
Where \sys's P99 finally rises, its own tokenization still takes only
\qtyrange{2.4}{15.2}{\milli\second}, and the bottom row of
Figure~\ref{fig:dynamo-fleet} shows it nearly flat across every load.  At
one million tokens and 46.25 requests/s, the largest load \sys sustains,
its tokenize stage takes \qty{11.5}{\milli\second} of a
\qty{454}{\milli\second} median \gls{ttft}, while gigatoken takes
\qty{80}{\milli\second} ($6.9\times$ longer) and reaches a P99 $12\times$
\sys's.  One load step further,
\sys's tokenize stage grows by \qty{2.4}{\milli\second} while its P99 grows
by \qty{2.1}{\second}.  The limit is the frontend's request path, which
parses multi-megabyte request bodies, renders templates, hashes token
blocks, and sends token IDs to the workers.  A serialized stage in this
path, rather than tokenization, bounds the frontend.  A control run reproduces the limit without any
tokenization.  We replaced the tokenizer with a host-side stub that returns
prefix-stable IDs hashed from the text, with no vocabulary and no merges, at
a per-request cost similar to \sys's (\qtyrange{1.4}{17.6}{\milli\second}).
With the stub, the frontend sustains 444, 111, and 37 requests/s, within one
load step of \sys at every length and never above it.  One frontend with \sys
therefore feeds as many GPUs as Dynamo's request path allows: the traffic of
56, 14, and 5 GPUs at the three lengths, against 32, 8, and 2 with gigatoken.

\takeaway{Within the \qty{2}{\second} P99 target, one Dynamo frontend
sustains the most load with \sys at every context length, $1.75$--$2.5\times$
the load it sustains with gigatoken and $28\times$ the load with Dynamo's
default tokenizer at 121k and 512k tokens.  With \sys, Dynamo's own request
path, not tokenization, becomes the limit.}

\subsection{Does it hold on a host shared with coding agents?}
\label{sec:eval-agentic}

\begin{figure*}[t]
  \centering
  \includegraphics[width=\textwidth]{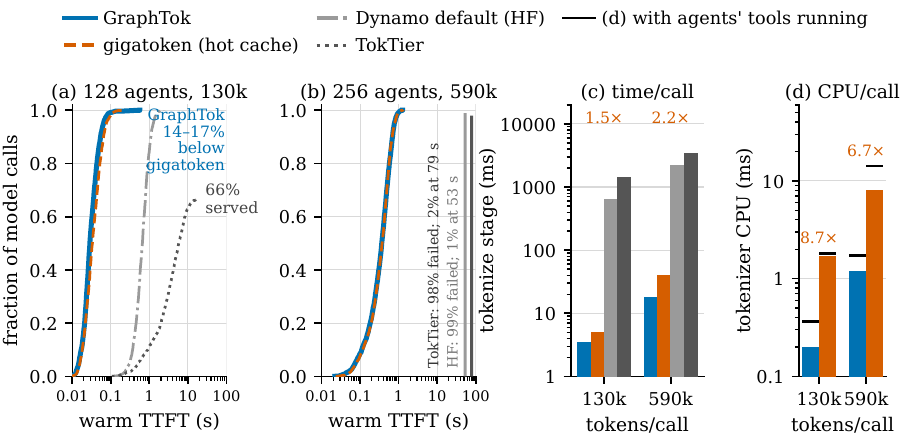}
  \caption{Coding agents sharing the frontend's host: Dynamo's frontend in front of simulated GPUs, and 128 or 256 agents that each work
  through a long conversation of real SWE-smith trajectories and run real
  tool calls on the same host.  (a, b)~CDF of the time to first token of
  every warm model call.  A curve that ends below one shows the fraction of
  calls served at all.  In (b), Hugging Face and TokTier serve only 1--2\%
  of calls, so each is drawn as a bar at the median time to first token of
  the calls it served, as tall as the share of calls it failed.  (c)~Dynamo's tokenize stage per call (mean).
  (d)~Host CPU the tokenizer itself spends per call, from replaying the
  runs' prompts through each library on the frontend's CPUs (bars with
  nothing else running, ticks with the 128 agents' tool calls running).  Labels in
  (c) and (d) give gigatoken's cost relative to \sys's.}
  \label{fig:agentic-host}
   \Description{Two CDF panels and one bar panel.  Panel a, 128 agents with
    130k-token calls: GraphTok's curve is leftmost and below gigatoken at every
    percentile, while Hugging Face and TokTier serve only a fraction of calls.
    Panel b, 256 agents with 590k-token calls: GraphTok and gigatoken overlap
    and serve every call, while bars at their served calls' median show that Hugging Face and
    TokTier fail 99\% and 98\% of calls.
    Panel c: the tokenize stage per call, lowest for GraphTok.  Panel d: the
    tokenizer's own host CPU per call, lowest for GraphTok.}
\end{figure*}

The fleet experiment gives the frontend a quiet host, whereas an agent
platform packs many agent sandboxes onto each machine.  DeepSeek's sandbox
infrastructure reports per-node peaks of 524 microVMs and 1,048 containers
on 384 hardware threads~\cite{huang2026dsec}, and every build, test, and
search an agent runs uses the same CPUs as everything else on that host.
We therefore place the same Dynamo frontend (with its simulated GPUs) on our
96-thread host together with 128 coding agents, the microVM density scaled
to this host, and in a second run with 256 agents, the container density,
and longer contexts.

Each agent is its own process and works through one long conversation built
from real SWE-smith coding-agent trajectories~\cite{yang2025swesmith}.
Whole earlier tasks form its history, and every model call re-sends the
entire conversation plus the next real turn of the trajectory, a median of
about 400 new tokens.  Every call is thus a re-tokenization of a long and
almost entirely repeated prompt, the pattern TraceLab measures for coding
agents (a median call of 119k cached-prefix tokens with 875 new
ones)~\cite{zhu2026tracelab}.  After each call, the agent waits out the
call's duration, drawn from the Aries production agentic
trace~\cite{kondrashov2026aries}, and then executes one real tool call on
the host: a file edit, a code search, a compile, or a test run with BLAS
bursts.  Histories are drawn so that calls centre on TraceLab's median in
the first run (130k tokens) and spread up to the 1M-token context limit in
the second (590k median).  Every tokenizer sees the same sessions, timings,
and tool calls during a six-minute run in which the agents start at random
within the first 30~s.  We exclude the first call of each agent, which
resumes its conversation with nothing cached.  With unrelated jobs also
running, the host stays 95--98\% busy throughout.

\smartparagraph{\sys is the fastest even where caching favours the CPU.}
At TraceLab's median length (Figure~\ref{fig:agentic-host}(a)), \sys has
the lowest time to first token at every percentile, with \qty{29}{},
\qty{54}{}, and \qty{93}{\milli\second} at the median, 90th, and 99th
percentiles.  This is 14--17\% below gigatoken (\qty{33}{}, \qty{65}{}, and
\qty{109}{\milli\second}), even though re-sent prefixes are exactly the case
gigatoken's pre-token cache is built for.  Dynamo's default tokenizer
is $23\times$ slower at the median (\qty{662}{\milli\second}), and TokTier
serves only 66\% of calls, with a median of \qty{3.3}{\second}.  Dynamo's
tokenize stage takes \qty{3.4}{\milli\second} per call with \sys,
$1.5\times$ that with gigatoken, and \qty{632}{\milli\second} with the
default (Figure~\ref{fig:agentic-host}(c)).

\smartparagraph{Longer contexts break the tokenizers that ship.}
With 256 agents and calls of 590k tokens (Figure~\ref{fig:agentic-host}(b)),
Dynamo's default tokenizer and TokTier serve only 1\% and 2\% of calls,
because a CPU tokenizer that needs seconds per call cannot keep up on a
host whose CPUs the agents keep busy.  \sys and gigatoken both serve every
call, with medians within 2\% of each other (\qty{365}{} and
\qty{373}{\milli\second}).  At this length, templating megabytes of
conversation and sending the token IDs to the workers dominate the
frontend's path, although \sys still tokenizes $2.2\times$ faster
(\qty{17.7}{} against \qty{38.9}{\milli\second}).

\smartparagraph{\sys takes the tokenizer off the host's CPUs.}
On such a host, a CPU tokenizer and the agents compete for the same cores.
Every core that gigatoken occupies is unavailable to the agents' compiles,
tests, and searches, whereas \sys tokenizes on the GPU and leaves those cores
to the agents.  Time to first token does not reveal this cost.  To isolate the tokenizer's own cost, we replay warm prompts
of both runs (1,024 calls averaging 128k tokens and 512 averaging 569k,
rendered with the model's chat template) through the two libraries, one call
at a time on the frontend's CPUs, with the \sys call configured as in
Section~\ref{sec:eval-toktier} (every input on the GPU, blocking host wait),
and measure the process's CPU time over all of its threads.  Both libraries
return identical token IDs for every prompt, and the reference's IDs on a
sample.  \sys spends \qty{0.19}{\milli\second} of host CPU per 130k-token
call and \qty{1.18}{\milli\second} per 590k-token call, $8.7\times$ and
$6.7\times$ less than gigatoken's \qty{1.68}{} and \qty{7.90}{\milli\second}
(Figure~\ref{fig:agentic-host}(d)).  With the 128 agents' tool calls running
on the host, the ratios are $5.0\times$ and $8.2\times$, as gigatoken's cost
per 590k-token call nearly doubles to \qty{14.1}{\milli\second} while
\sys's rises to \qty{1.7}{\milli\second}.  The call also returns sooner: in
\qty{0.69}{} against \qty{1.68}{\milli\second} at 130k tokens and
\qty{2.74}{} against \qty{7.62}{\milli\second} at 590k with nothing else running
($2.4\times$ and $2.8\times$), and $2.2\times$ and $4.2\times$ faster with
the agents' tools running.

The mechanism is the one Table~\ref{tab:dispatch} shows for short requests.
The dispatcher's host work consists of staging the text and copying the IDs
back, and its thread sleeps while the GPU works, whereas a CPU tokenizer's
work grows with every token it reads, cached or not.  The saving is visible
in the Dynamo runs themselves, where the whole frontend spends 13\% less CPU
per call with \sys at 130k tokens (\qty{23.8}{} against
\qty{27.5}{\milli\second}); the remainder is request parsing, templating,
and transport, which both tokenizers share.  At these per-call costs, every
1,000 calls/s of 130k-token prompts keeps 1.7--1.8 host cores busy
tokenizing with gigatoken and 0.2--0.4 with \sys.  At 590k tokens,
gigatoken takes 8--14 cores from the agents and \sys takes 1.2--1.7.

\smartparagraph{The work moves to the GPU without slowing the model.}
Moving tokenization off the host places it on a GPU, so we measure what it
costs a model that shares that GPU.  vLLM serves Qwen3-32B on the H100
under a heavy load (16 concurrent streams, each request prefilling a
never-repeated 2k-token prompt and generating 128 tokens), while a separate
process, as a serving frontend would be, tokenizes the agentic prompts with
\sys on the same GPU at the two runs' rates.  Its calls occupy at most
1.3\% of the GPU's time at 5.5~calls/s of 130k-token prompts and 6.7\% at
10.8~calls/s of 590k-token prompts, and use 3--5\,GB of the GPU's 94\,GB of
memory.  The model does not slow down measurably.  Over nine alternating
one-minute windows, mean throughput with and without the tokenizer differs
by 1.1\%, against a 2.9\% window-to-window spread in the baseline alone.
With and without the tokenizer, throughput is 369--372 versus
369--381~tokens/s, median inter-token latency
\qtyrange{17.2}{17.6}{\milli\second} versus
\qtyrange{16.9}{17.3}{\milli\second}, and median time to first token
(here, prefill) \qtyrange{267}{281}{\milli\second} versus
\qtyrange{263}{353}{\milli\second}; in each metric the range without the
tokenizer is at least as wide as the range with it.
The same holds inside vLLM, where \sys shares the model's process: prefill
and inter-token latency stay within
2.5\% and 1.5\% of the runs with gigatoken at every length.  It is the
tokenizer that yields.  Queued behind the model's kernels, the same call
takes \qty{2.3}{} and \qty{4.7}{\milli\second} in median instead of the
\qty{0.69}{} and \qty{2.74}{\milli\second} of the idle-GPU replay, with its
host CPU time unchanged.  At these rates, \sys needs only a few percent of
one inference GPU, and a frontend needs no GPU of its own.

\begin{figure}[t]
  \centering
  \includegraphics[width=\columnwidth]{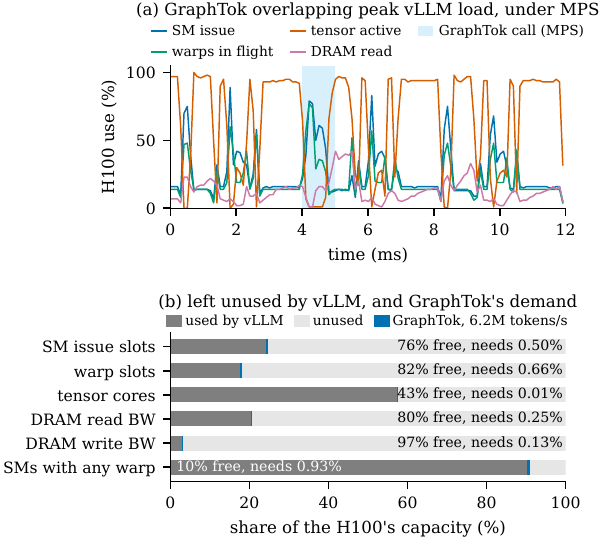}
  \caption{\sys overlaps the H100 capacity that steady-state vLLM inference
  leaves stranded.
  (a)~Twelve milliseconds of the Nsight Systems hardware counters (10\,kHz)
  captured on one clock while vLLM (Qwen3-32B) serves its peak steady-state
  load (32 streams) and \sys\ tokenizes 590k-token agentic prompts on the same
  H100, both under MPS; the shaded band is \sys's real 1.0\,ms call, and because
  MPS co-schedules the two contexts the model's tensor activity stays up through
  the burst while \sys\ adds issue, warp, and DRAM demand.  (b)~Mean share of
  each resource that vLLM uses and leaves unused at that peak load, against the
  time-averaged demand of \sys\ tokenizing 10.8~calls/s of 569k-token prompts
  (6.2M tokens/s).}
  \label{fig:stranded}
  \Description{Panel (a) plots SM issue, warps in flight, tensor activity and
  DRAM read bandwidth over 12 milliseconds captured on one clock while GraphTok
  and vLLM run concurrently under MPS; a shaded band marks GraphTok's real 1.0
  millisecond call, during which the model's tensor activity stays elevated
  while SM issue and DRAM rise.  Panel (b) shows horizontal bars per resource:
  at its peak load vLLM uses 24 percent of issue slots, 18 percent of warp
  slots, 57 percent of tensor cores, 20 percent of DRAM read and 3 percent of
  DRAM write bandwidth, and 90 percent of SMs have at least one warp; 76, 82,
  43, and 80 percent of the first four are left free.  GraphTok needs 0.50,
  0.66, 0.01, 0.25, 0.13 and 0.93 percent of these resources.}
\end{figure}

\smartparagraph{Inference leaves the resources \sys needs idle.}
Steady-state LLM inference does not saturate the GPU: decoding is bound by
memory bandwidth and leaves compute idle, and prefill and decode stress
different units~\cite{patel2024splitwise,zhu2025nanoflow,kamath2025podattention,siavashi2026dissecting}.
Figure~\ref{fig:stranded} shows this on the H100 at 32 streams, the highest
load vLLM sustains with the workload above before throughput falls and
requests queue.  Even there, vLLM issues instructions in 24\% of the SMs'
issue slots, occupies 18\% of the warp slots, and uses 57\% of the tensor
cores and 20\% of the DRAM read bandwidth, leaving 76\%, 82\%, 43\%, and
80\% of them unused.  \sys, tokenizing 6.2M tokens/s of agentic prompts,
needs 0.50\% of the issue slots, 0.66\% of the warp slots, 0.25\% of the
DRAM read bandwidth, and no tensor cores, at least $10\times$ less than the
headroom on every resource it uses.  Panel (a) shows this on a shared
clock, with vLLM and \sys\ running concurrently on the same H100.  Because
MPS or a fine-grained scheduler lets the two contexts share SMs rather than
time-slice, the model keeps its tensor activity through \sys's call while
\sys\ adds only the issue, warp, and DRAM demand it needs; the model
sustains \qty{434}{tokens/s} against \qty{436}{tokens/s} alone, so the
overlap does not measurably slow it.

\takeaway{On a host shared with coding agents, \sys matches or beats the fastest CPU
tokenizer in that tokenizer's most favourable regime, eliminates the failures of the tokenizers that Dynamo and TokTier
ship, spends $8.7\times$ and $6.7\times$ less host CPU per call than
gigatoken, and fits within the small share of one inference GPU that the model
leaves idle.}

\subsection{Where does the time go, and how much does graph replay save?}
\label{sec:eval-execution}

\begin{figure}[t]
  \centering
  \includegraphics[width=\columnwidth]{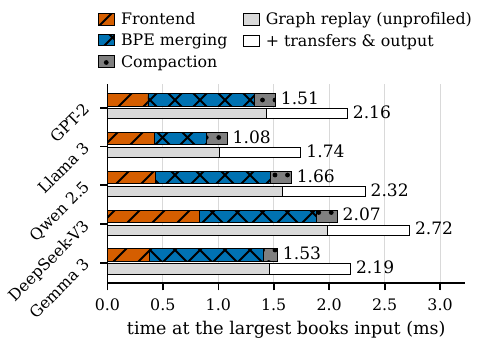}
  \caption{Stage profile at the largest books input: profiled GPU time
  per stage (upper bars) and the complete call as graph replay plus transfers
  (lower bars).}
  \label{fig:stage-breakdown}
  \Description{Five pairs of horizontal bars for GPT-2, Llama 3, Qwen 2.5,
  DeepSeek-V3, and Gemma 3 on the books corpus.  In each profiled bar, BPE
  merging is the largest stage and compaction the smallest, while frontend
  cost varies by tokenizer.  Profiled totals range from 1.08 to 2.07
  milliseconds.  The paired bars show complete calls of 1.74 to 2.72
  milliseconds, of which graph replay is the larger part.}
\end{figure}

\smartparagraph{Pipeline breakdown.}
Having established that the complete call is fast, we examine which of its
stages dominate.  Figure~\ref{fig:stage-breakdown} profiles the captured
pipeline on books at one million tokens.  \gls{bpe} merging is the largest
stage for all five tokenizers and output compaction is consistently the
smallest, while the frontend's cost varies with the normalization and
pre-tokenization rules of each family.  Pre-tokenization, the stage that
prior GPU tokenizers omitted because they considered it too expensive, is
therefore not where the time goes.  Stage times are medians of 20 profiled
replays, and the lower bars time graph replay with CUDA events in an
unprofiled run.  The profiled stages sum to
\qtyrange{1.08}{2.07}{\milli\second}, 4--7\% above the unprofiled graph
replay of the same input, because the sum counts concurrently running graph
nodes in full.  The complete host-returning call takes
\qtyrange{1.74}{2.72}{\milli\second}, and the remainder over graph replay
 is transfer and output materialization.  The profile drives \sys's
standalone binary; the build-to-build speed changes do not alter the
relative stage shares, which are the point of the breakdown.

\smartparagraph{Graph replay.}
The remaining question is how much the captured graph itself contributes,
as opposed to the parallel kernels it contains.  To answer it, we run the
same GPT-2 pipeline and host-returning interface under graph replay and
under individual kernel launches.  For each of seven books inputs from 128
to 100,000 tokens, we collect 100 measurements in alternating ten-sample
blocks after five warmups and verify the output against tiktoken.  Both
modes reuse the capacity-specific launch plan prepared during warmup, so
the difference isolates dispatch rather than repeated device queries or
grid calculations.  Graph replay reduces the median complete-call latency
by 21--23\% from 128 to 2k tokens and by 20\% at 8k, and the reduction
falls to 12\% at both 32k and 100k.

\takeaway{\gls{bpe} merging, not pre-tokenization, is the largest stage for all five
tokenizers. Graph replay saves 21--23\% of the complete call near the
short-input floor, where dispatch is a large share of the call, and still about
an eighth once useful GPU work dominates.}

\subsection{Limitations}

\sys implements five concrete tokenizer configurations rather than
arbitrary \texttt{tokenizer.json} files, and the serving integrations are
validated for Llama~3, Qwen~2.5, and Qwen~3.  This is a deliberate
exactness-first choice (Section~\ref{sec:design}): adding a new family
requires a new frontend predicate, and the generality of the result is
bounded by the five evaluated configurations.  All experiments use one H100
NVL, so none establishes hardware scaling.

The two vLLM serving experiments (Sections~\ref{sec:eval-integration}
and~\ref{sec:eval-burst}) serve one model at a time on one GPU, where the tokenizer is lightly loaded, so
neither measures a saturated multi-GPU engine.  The CPU-route crossover is
tuned on never-repeated text, so on a repeated short document the router
can keep the graph where a warm CPU engine would be faster (by up to
$2.4\times$ at batch~1 in Figure~\ref{fig:batch-scaling}); the router's
content adaptation narrowly covers this case at the measured limits, and
the cost is the graph's fixed floor on short repeated documents that a
warm CPU engine serves faster.  The Dynamo experiment simulates its GPUs
(Section~\ref{sec:eval-fleet}), so it measures the frontend's capacity
rather than end-to-end \gls{ttft}, and it runs with Dynamo's prefix
tokenizer cache off, which is the default; in multi-turn traffic whose
history ends at a special token, that cache would shorten the input of
every tokenizer.  The agentic-host experiment shares that simulation and
runs each tokenizer once on a host shared with unrelated jobs; its
tokenizer CPU per call is measured by replay outside Dynamo.  The
GPU-sharing measurement covers one model, one load, and two tokenization
rates.

\section{Related Work}
\label{sec:related}
\subsection{Tokenization acceleration}

Making tokenization faster on the CPU is the most direct alternative to our
approach, and several systems pursue it. Fast WordPiece fuses
pre-tokenization and vocabulary matching for WordPiece, a different
algorithm from the byte-level \gls{bpe} considered
here~\cite{song2021fastwordpiece}. For \gls{bpe}, Fastokens adds CPU
parallelism and shared caches~\cite{kejzman2026fastokens}, and gigatoken
combines SIMD processing, parallel batches, and token
caches~\cite{roed2026gigatoken}. LoPT takes a
different route to parallelism. It partitions long inputs into overlapping
regions, tokenizes them independently, and reconciles their outputs
exactly~\cite{shao2025lopt}. All of these approaches keep the token IDs on
the host. \sys produces them on the accelerator and supports
device-resident output, although the evaluated vLLM adapter returns the IDs
to the host. The device path serves CPU-free engines such as
Blink~\cite{siavashi2026blink} (Section~\ref{sec:design}).

GPU tokenization has been explored for two families of tokenizers. RAPIDS
provides a batched WordPiece tokenizer for BERT
workloads~\cite{rapids2020gst}, and two systems implement byte-level
\gls{bpe} kernels. BlockBPE removes the model's regular-expression
 pre-tokenizer~\cite{you2025blockbpe}, and GPUTOK moves GPT-2 merging to the
GPU but keeps byte encoding on the host and cannot be captured as a
graph~\cite{kadamba2026gputok}. \sys instead preserves the family-specific
frontend and captures the complete pipeline in one preallocated graph.
Section~\ref{sec:eval-gputok} reports how GPUTOK's output compares with the
reference tokenization.
TokTier~\cite{zhang2026toktier} is concurrent, independent work on exact
GPU tokenization. Neither system builds on the other, and we compare against
its released package in Section~\ref{sec:eval-toktier}.

\subsection{Reusing tokenization and prompt state}

Rather than tokenizing faster, a system can try to tokenize less. Caching
token IDs can bypass tokenization for text that recurs, but arbitrary
fragments are unsafe reuse units, because \gls{bpe} merges and pre-token
boundaries may change at a join. NVIDIA Dynamo therefore reuses token IDs only
at safe boundaries (Section~\ref{sec:motivation})~\cite{nvidia2026dynamotokenizercache}
and must still encode after a miss or an edit.

Exact incremental and streaming algorithms avoid re-encoding an unchanged
prefix in a more principled way. Jiang and Gong maintain the \gls{bpe}
tokenization as the text grows, and Mamouras et al.\ bound the lookahead
needed to emit a canonical token stream~\cite{jiang2026incrementalbpe,%
mamouras2026streamingbpe}. Both assume reusable state or a growing stream.
These approaches are complementary to \sys. Reuse reduces the amount of
work, whereas \sys accelerates the encoding of new text and of whatever
remainder still requires tokenization.

Most \emph{prompt caching} reuses model state after tokenization.
Content-based prefix caches such as SGLang's RadixAttention and vLLM's
automatic prefix caching match token prefixes and retain their \gls{kv}
state~\cite{zheng2024sglang,vllm2026prefixcache}. These interfaces require
the token IDs before matching can begin, so their reuse of model state
leaves exactly the tokenization cost that \sys addresses.

\section{Conclusion}
\label{sec:conclusion}
\sys shows that GPU tokenization need not sacrifice exactness. By evaluating
pre-tokenization rules at every character in parallel, merging within warps
in the reference order, and running the whole encode as one capacity-sized
graph, it reproduces the reference token IDs of five production tokenizers in
every evaluated cell. Above the CPU-route crossover, it is the fastest engine
we measure on the standalone, batched, and TokTier single-request protocols.

The latency benefit is real, and where it lands is predictable. In unmodified vLLM it pays off once
prefix reuse has removed enough prefill to expose the frontend, and in that
regime fast CPU tokenizers come within 3--18\% of \sys on a single
uncontended request. Where one frontend tokenizes for a fleet, the benefit
becomes capacity, and with \sys Dynamo's own request path, not tokenization,
becomes the limit. As caching and faster prefill shrink the GPU's share of a
request, the fixed host stages that remain determine its latency, and each
will have to be moved or removed in turn. The next step is an engine that
consumes device-resident token IDs directly, as CPU-free systems such as
Blink~\cite{siavashi2026blink} already do for the rest of the stack, which
removes the remaining transfer and synchronization from the request path.

\bibliographystyle{mlsys2025}
\bibliography{refs}

\end{document}